\documentclass[twocolumn,aps,epsfig,nofootinbib]{revtex4}

\usepackage{graphicx}
\usepackage{epstopdf}
\usepackage{latexsym}
\usepackage{amssymb}
\usepackage{amsmath}
\usepackage{mathrsfs}
\usepackage{braket}
\usepackage{color}
\usepackage[center]{subfigure}

\begin{document}

 \newcommand{\bq}{\begin{equation}}
 \newcommand{\eq}{\end{equation}}
 \newcommand{\bqn}{\begin{eqnarray}}
 \newcommand{\eqn}{\end{eqnarray}}
 \newcommand{\nb}{\nonumber}
 \newcommand{\lb}{\label}
\newcommand{\PRL}{Phys. Rev. Lett.}
\newcommand{\PL}{Phys. Lett.}
\newcommand{\PR}{Phys. Rev.}
\newcommand{\CQG}{Class. Quantum Grav.}
\title{Non-Gaussianity of a single scalar field   in general covariant Ho\v{r}ava-Lifshitz gravity}

\author{Yongqing Huang}
\email{yongqing_huang@baylor.edu}

\author{Anzhong Wang}
\email{anzhong_wang@baylor.edu}

\affiliation{ 
Institute for Advanced Physics $\&$ Mathematics, Zhejiang University of
Technology, Hangzhou 310032,  China  \\
GCAP-CASPER, Physics Department, Baylor
University, Waco, TX 76798-7316, USA  }

\date{\today}

\begin{abstract}
In this paper, we study  non-Gaussianity generated by a single scalar field in slow-roll inflation in the 
framework of the non-relativistic general covariant Ho\v{r}ava-Lifshitz theory of gravity with the projectability 
condition and an arbitrary coupling constant $\lambda$, where $\lambda$ characterizes the deviation of the
theory from general relativity (GR) in the infrared. We find that the leading effect of self-interaction, in contrary to the 
case of minimal scenario of GR, is in general of the order $\hat{\alpha}_{n} \epsilon^{3/2}$, 
where $\epsilon$ is a slow-roll parameter, and $\hat{\alpha}_{n}\; (n = 3, 5)$ are the dimensionless coupling 
coefficients of the six-order operators of the Lifshitz scalar,  and have no contributions to  power spectra and 
indices of both scalar and tensor.  The bispectrum, comparing with the standard one given in 
GR, is enhanced, and gives rise to  a large value of the nonlinearity parameter $f_{\text{NL}}$.
We study how the modified dispersion relation with high order moment terms  affects the evaluation of 
the mode function and in turn the bispectrum, and show explicitly that the mode function takes various 
asymptotic forms during different periods of its evolution. In particular, we find that it is  in general of 
superpositions of oscillatory functions, instead of  plane waves like in the minimal  scenario of GR.
This results in a large enhancement of the folded shape in the bispectrum.
\end{abstract}

\pacs{ 98.80.Cq, 98.80.-k, 04.50.-h}

\maketitle
\section{Introduction}
\renewcommand{\theequation}{1.\arabic{equation}} \setcounter{equation}{0}

The   Ho\v{r}ava-Lifshitz (HL) theory of   quantum gravity, proposed recently by Ho\v{r}ava \cite{Horava}, motivated by the Lifshitz 
scalar field theory in solid state physics \cite{Lifshitz}, has attracted a great deal of attention, due to
its several remarkable features in cosmology as well as some challenging questions, such as instability, ghost, strong coupling,
and different speeds \cite{reviews}.  To resolve  these questions, various  models have been proposed, along two fundamentally 
different lines, one with the projectability condition   \cite{SVW,HMT,WW,dS,HW},
\bq
\lb{0.1}
N = N(t),
\eq
 and the other without it \cite{BPS,ZWWS},  where 
$N$ denotes the lapse function in the Arnowitt, Deser and Misner decompositions  \cite{ADM}. 
In particular,  Ho\v{r}ava and Melby-Thompson (HMT)  proposed to enlarge the foliation-preserving diffeomorphisms of the original model, 
\bq
\lb{0.2}
\delta{t} = - f(t),\; \;\; \delta{x}^{i}  =  - \zeta^{i}(t, {\bf x}),
\eq
often denoted by Diff($M, \; {\cal{F}}$), to include   a local U(1)  symmetry, so that the reformulated  theory  has the symmetry  \cite{HMT},
\bq
\lb{symmetry}
 U(1) \ltimes {\mbox{Diff}}(M, \; {\cal{F}}).
 \eq
With such an enlarged symmetry, the spin-0 gravitons, which appear in the original model of the HL theory \cite{Horava}, are eliminated
\cite{HMT,WW}, and as a result, all the problems related to them,  including instability, ghost, and strong coupling, are resolved automatically.
This was initially done with $\lambda = 1$, where $\lambda$ characterizes the deviation of the
theory from general relativity (GR) in the infrared, as one can see from  Eqs.(\ref{1.1}) and (\ref{1.2}) given below. Soon, it was generalized
to the case with any $\lambda$ \cite{dS}, in which it was  shown that the spin-0 gravitons are also eliminated   \cite{HW,dS}, so that the above 
mentioned problems are resolved in the gravitational sector.
In the matter sector, the strong coupling problem, first noted  in \cite{HW},  can be solved by introducing a mass $M_{*}$ so that
$M_{*} < \Lambda_{\omega}$, where $M_{*}$ denotes the suppression energy of high order operators, and $\Lambda_{\omega}$
 the would-be energy scale, above which matter becomes strongly coupled
\cite{LWWZ}, similar to the non-projectability case without the enlarged symmetry  \cite{BPSb}. The consistence of this model with solar system 
tests was investigated recently \cite{LMW}, and found that it is consistent with all such tests, provided that the gauge field and Newtonian prepotential 
are part of the metric, as proposed originally by HMT \cite{HMT}. 

In this paper, we shall work within the HMT framework of the HL theory with the projectability condition \cite{HMT,WW,dS,HW}. 
For the current status of  the models proposed in  
\cite{SVW,BPS,ZWWS}, we refer readers  to \cite{LMW}. 

Since the HL theory differs from GR significantly in high energies, in this paper we study another important issue
- the primordial non-Gaussianity in   the HL theory. Such a problem was studied previously, one without the projectability condition
\cite{Gao}, and the other with it \cite{Shinji-II} but in the curvaton scenario \cite{Curvaton}, and many interesting results were obtained.
However, our study presented in this paper are different from these at least in two aspects: (i) we shall study the problem
in the framework of the non-relativistic general covariant Ho\v{r}ava-Lifshitz theory of gravity with the projectability 
condition \cite{HMT,WW,dS,HW}, as mentioned above, in which the degree of gauge freedom is the same as that in GR; and (ii) we shall investigate the problem in the inflationary
scenario  for a single scalar field with slow-roll conditions \footnote{To solve the horizon problem and obtain almost scale-invariant perturbations
in the HL theory, inflation is not necessary, as first noted in \cite{ShinjiInflation}. But, to solve othe problems, such as the relics of topological defects,
including monopoles, it is still needed \cite{Inflation}.}. Because of these differences,  our results are also significantly different from
theirs.

To study non-Gauusianity, various techniques have been developed. In particular, the in-in formalism, developed initially
by Schwinger some decades ago \cite{Schwinger},  becomes    standard, after it was first explored  by Madalcena in his pioneer work 
\cite{Madalcena} in calculating the high-order correlators for cosmological perturbations, and then further developed  by 
Weinberg in his seminal paper \cite{Weinberg05}. For reviews, see \cite{Koyama,Contractions}. 
In this paper, we will follow this approach. 

Non-Gauusianities have attracted lot of attention recently and been studied extensively in various models  \cite{Koyama,Contractions}, mainly because 
they could be well within the range of detection  of the current Planck satellite \cite{Planck} and the forthcoming experiments, such as  the  CMBPol mission \cite{CMBPol}. 
Among  various models, the one with a canonically coupled single scalar field in the framework of GR, though the simplest, gives elegant predictions and fits extremely  
well with current observations \cite{COBE}. It predicts that the quantum fluctuations, which are responsible for generating the primordial perturbations and in turn the 
anisotropies in CMB and inhomogeneity in the large scale structure, are largely Gaussian. The effect of the non-linearity in the primordial perturbations is  an order 
smaller than that of the power spectrum and is beyond our current detectabilities.

However, Holman and Tolley  recently  argued that,  while a change in high energy physics during or before inflation gives mild modifications to the power spectrum
 and index, the non-Gaussianity, evaluated by higher order correlation functions of the perturbations, is much more sensitive to the new physics
 in the ultraviolet (UV) \cite{Holman}. Therefore,  it is very interesting to study the non-Gaussianity in the scalar primordial perturbations of the HL theory, 
 where the dispersion relations are quite different from the standard one. 
 
The rest of the paper is organized as follows. In Section II we give a brief review of the  non-relativistic general covariant Ho\v{r}ava-Lifshitz theory of gravity with the projectability 
condition and an arbitrary coupling constant $\lambda$. The self-interaction Hamiltonian and the leading order terms are analyzed in Section III. Section IV discusses the modified 
dispersion relation and its effect on the mode function and the shapes of bispectrum. We also compare our results with those presented in  \cite{Shinji-II}. Finally, in Section V
we summarize our results and the assumptions made along the way of analysis. Three appendices are also included, where in Appendix A the linear perturbations and the corresponding 
field equations are presented, while in Appendix B, the cubic part of the total action is calculated, and given explicitly. In Appendix C, the matching across the boundaries among the 
three different regions of the time revolution of the mode function [cf. Fig. \ref{fig0}] are given.

\section{General covariant HL gravity with projectability condition}
\renewcommand{\theequation}{2.\arabic{equation}} \setcounter{equation}{0}

 The action of the general covariant HL theory of gravity with the projectability condition   can be written as \cite{HMT,WW,dS,HW},
\bqn
\lb{1.1}
S &=& \frac{1}{16\pi G}\int dt d^{3}x N \sqrt{g} \Big({\cal{L}}_{K} -
{\cal{L}}_{{V}} +  {\cal{L}}_{{\varphi}} +  {\cal{L}}_{{A}} +  {\cal{L}}_{{\lambda}}\Big) \nb\\
& & ~~~~ + \int dt d^{3}x N \sqrt{g} {\cal{L}}_{M},
\eqn
where $g={\rm det}(g_{ij})$, $G$ is the Newtonian constant, and
\bqn
\lb{1.2}
{\cal{L}}_{K} &=& K_{ij}K^{ij} -   \lambda K^{2},\nb\\
{\cal{L}}_{{V}} &=& \zeta^{2}g_{0}  + g_{1} R + \frac{1}{\zeta^{2}}
\left(g_{2}R^{2} +  g_{3}  R_{ij}R^{ij}\right)\nb\\
& & + \frac{1}{\zeta^{4}} \left(g_{4}R^{3} +  g_{5}  R\;
R_{ij}R^{ij}
+   g_{6}  R^{i}_{j} R^{j}_{k} R^{k}_{i} \right)\nb\\
& & + \frac{1}{\zeta^{4}} \left[g_{7}R\Delta R +  g_{8}
\left(\nabla_{i}R_{jk}\right)
\left(\nabla^{i}R^{jk}\right)\right],\nb\\
{\cal{L}}_{\varphi} &=&\varphi {\cal{G}}^{ij} \Big(2K_{ij} + \nabla_{i}\nabla_{j}\varphi\Big),\nb\\
{\cal{L}}_{A} &=&\frac{A}{N}\Big(2\Lambda_{g} - R\Big),\nb\\
{\cal{L}}_{\lambda} &=& \big(1-\lambda\big)\Big[\big(\Delta\varphi\big)^{2} + 2 K \Delta\varphi\Big].
\eqn
Here $\Delta \equiv g^{ij}\nabla_{i}\nabla_{j}$,  
$\Lambda_{g}$ is a    coupling constant,  the
Ricci and Riemann tensors $R_{ij}$ and $R^{i}_{jkl}$  all refer to the 3-metric $g_{ij}$, and
\bqn
\lb{1.3}
K_{ij} &=& \frac{1}{2N}\left(- \dot{g}_{ij} + \nabla_{i}N_{j} +
\nabla_{j}N_{i}\right),\nb\\
{\cal{G}}_{ij} &=& R_{ij} - \frac{1}{2}g_{ij}R + \Lambda_{g} g_{ij}.
\eqn
 The coupling constants $ g_{s}\, (s=0, 1, 2,\dots 8)$ are all dimensionless, and
 \bq
 \lb{lambda}
 \Lambda = \frac{g_0}{32\pi G},
 \eq
 is the cosmological constant. The relativistic limit in the IR, on the other hand, 
 requires, 
 \bq
 \lb{NC}
 g_{1} = -1.
 \eq
${\cal{L}}_{{M}}$ is the Lagrangian of matter fields, and for a scalar field $\chi$, it is given by \cite{WWM,Inflation}, 
\bqn
\lb{1.6a}
{\cal{L}}_{M} &=& {\cal{L}}^{(0)}_{\chi} + {\cal{L}}^{(A,\varphi)}_{\chi},\nb\\
{\cal{L}}^{(0)}_{\chi} &=& \frac{f(\lambda)}{2N^{2}}\Big(\dot{\chi} - N^{i}\nabla_{i}\chi\Big)^{2} - {\cal{V}},\nb\\
{\cal V}  &=& V\left(\chi\right) + \left(\frac{1}{2}+V_{1}\left(\chi\right)\right) (\nabla\chi)^2
+  V_{2}\left(\chi\right){\cal{P}}_{1}^{2}\nb\\
& & +  V_{3}\left(\chi\right){\cal{P}}_{1}^{3}  +
V_{4}\left(\chi\right){\cal{P}}_{2} + V_{5}\left(\chi\right)(\nabla\chi)^2{\cal{P}}_{2}\nb\\
& & 
+ V_{6}(\chi) {{\cal P}}_{1} {\cal{P}}_{2}, \\
\lb{1.6b}
{\cal{L}}^{(A,\varphi)}_{\chi} &=& \frac{A - {\cal{A}}}{N}  \Big[c_{1}\left(\chi\right)\Delta\chi + c_{2}\left(\chi\right)\big(\nabla\chi\big)^{2}\Big]\nb\\
&&  - \frac{f}{N}\Big(\dot{\chi} - N^{i}\nabla_{i}\chi\Big)\big(\nabla^{k}\varphi\big)\big( \nabla_{k}\chi\big)\nb\\
& & + \frac{f}{2}\Big[\big(\nabla^{k}\varphi\big)\big(\nabla_{k}\chi\big)\Big]^{2},
 \eqn
 with $c_1\left(\chi\right), c_2\left(\chi\right), V(\chi)$ and $V_{n}(\chi)$ being arbitrary functions of $\chi$, and
\bq
\lb{1.6c}
{\cal{P}}_{n} \equiv \Delta^{n}\chi.
\eq
 
 For detail, we refer readers to \cite{HW,Inflation}. In the following we shall use directly the symbols and conversions from these papers, without 
 further explanations.

\section{Non-Gaussianities} 
\renewcommand{\theequation}{3.\arabic{equation}} \setcounter{equation}{0}

The homogeneous and isotropic flat universe is described by,
\bq
\lb{eq1}
\hat{N} = a(\eta),\;\;   
\hat{g}_{ij} = a^{2}(\eta)\delta_{ij},\;\;   
\hat{N}_{i} = 0 = \hat\varphi,\;\; \hat{A} = \hat{A}(\eta), 
\eq
where   as in  \cite{Inflation},  we use symbols with bars to denote the quantities of the background in the ($t, x^i$)-coordinates, and the ones with hats
 to denote those in the ($\eta, x^i$)-coordinates, where the conformal time $\eta$ is defined as $\eta = \int{dt/a(t)}$. The linear perturbations, given by
 \bqn
\lb{eq2}
\delta{N} &=& a \phi,\;\;\; \delta{N}_{i} = a^{2}B_{,i},\nb\\
\delta{g}_{ij} &=& -2a^{2}\big(\psi \delta_{ij} - E_{,ij}\big),\nb\\
A &=& \hat{A} + a \delta{A},\;\;\; \varphi = \hat{\varphi} + \delta\varphi,\nb\\
\chi &=& \hat{\chi} + \delta\chi, 
\eqn
were studied in detail in \cite{Inflation}, and shown explicitly that a master equation exists for a single scalar field, with the gauge choice,
\bq
\lb{gauge}
\phi = 0 = E = \delta\varphi.
\eq
For the sake of reader's convenience, we summarize the main equations obtained in \cite{Inflation} in Appendix A of this paper. 

To generalize the above linear perturbations to the nonlinear case, in this paper we consider the perturbations, given by
\bqn
\lb{2.0a}
{g}_{ij} &=& a^{2} e^{2\zeta} \delta_{ij}, \;\;\; {N}_{i} = a^2B_{,i},\;\; N = \hat{N}, \nb\\
A &=& \hat{A} + a \delta{A},\;\; \varphi =  0,\;\; \chi = \hat{\chi} + \delta\chi. 
\eqn
Clearly,  to first order, they reduce to the ones given by Eq.(\ref{eq2}) if one identifies $\zeta$ as $\zeta = -\psi$. 
Substituting Eq.(\ref{2.0a}) into the total action (\ref{1.1}), we find that its cubic part is given by Eq.(\ref{C2}) in Appendix B.
From  the linear order constraint and field equations given by Eq.(\ref{b20.a})-(\ref{b20.b}) in Appendix A, we can express  
the terms  $B$ and $\delta A$ in terms of $\zeta$. Then,  substituting them into the cubic action (\ref{C2}), we find that
 it can be cast in the   schematic form, 
\bqn
\lb{2.1}
S = \int \sum^3_{m=0} \Big\{\partial^{2k} \zeta^m \cdot \partial^{2l}{\delta \chi}^{\left(3-m\right)}\Big\}.
\eqn

To find the leading order terms in the self-interaction, let us first note that \cite{Inflation}, 
\bqn
\lb{2.2a}
\delta\chi = h {\cal{R}} \propto \epsilon^{1/2} {M}_{\text{pl}}{\cal{R}},
\eqn
where 
${\cal{R}}$ is  the comoving curvature perturbations, defined explicitly in \cite{Inflation}.
From Eq.(\ref{b20.c}), we also have
\bq
\lb{2.2b}
\zeta \simeq -\psi = -4\pi G {c}_1  \delta \chi = - \frac{c_1 \delta\chi}{2M^2_{\text{pl}}}.
\eq
Assuming that  $c_1 \simeq M_{*} \ll M_{\text{pl}}$ \cite{Inflation}, we find that $\zeta \ll \delta\chi \ll {\cal{R}}$. This implies 
 that to find the leading order terms, it suffices to look for terms which are of cubic order of $\delta\chi$ [$m=0$ in (\ref{2.1})], 
 since all terms in lower orders of $\delta\chi$ are of higher orders of $\zeta$,
  hence further suppressed by factors of ${M_*}/{M_{\text{pl}}}$. With this as our guideline, it can be shown that
  only six terms are left for considerations. One from the part $S_{\chi}|^{\text{GR}}_{(3)}$ in Eq.(\ref{C3.2}), identified as
\bq
\lb{2.3a}
{V'''}(\chi)\delta\chi ^3.
\eq
However, since it is the third-order derivative term of the potential, one can  immediately ignore  it,  as the slow-roll conditions require
$|V'''| \ll 1$. The other five are from $S_{\chi}|^{\text{HL}}_{(3)}$ and are identified as, 
\bqn
\lb{2.3b}
&& \frac{V'_1}{a^2} \delta\chi\left(\partial_k\delta\chi\right)^2, \frac{V'_2}{a^4} \delta\chi \left(\partial^2\delta\chi\right)^2, 
 \frac{V''_4}{a^4}\left(\delta\chi\right)^2\left(\partial^4\delta\chi\right), \nb\\
&& \frac{V_3}{a^6} \left(\partial^2\delta\chi\right)^3,  \frac{V_5}{a^6} \left(\partial^4\delta\chi\right)\left(\partial_k\delta\chi\right)^2.
\eqn
Out of the five, three are proportional to derivatives of the coupling functions $V_1, V_2$, and $V_4$. They all appear in the 
linear  perturbations, and it was assumed that  their derivatives with respect to $\chi$ vanishes  \cite{Inflation}. To be consistent with it, in this paper we keep this assumption.
Hence, we are  finally  left only with two terms, that are proportional to  $V_3$ and $V_5$, 
\bqn
\lb{2.5}
&& \int \frac{d\eta d^3x}{a^2}h^3\Bigg\{V_3  \left(\partial^2{\cal{R}}\right)^3, V_5 \left(\partial^4{\cal{R}}\right)\left(\partial_k{\cal{R}}\right)^2\Bigg\}, ~~~~~
\eqn
where $h$ is defined in Eq.(\ref{b32}). In contrast to  the minimum scenario in GR, which predicts that the self-interaction should be of  the order 
of $\epsilon^2$, in the current case, the leading order is of $ \hat\alpha_n h^3  \sim \hat\alpha_{n}\epsilon^{3/2} $, where
$\hat\alpha_{n}$ are dimensionless parameters, defined by $V_{n} = \hat\alpha_{n}/M_{*}^4, \; (n = 3, 5)$. 

Despite the fact that these two terms are similar to  the $\alpha_2$ and $\alpha_3$ terms given in \cite{Shinji-II},
a key difference, however, exists. In  \cite{Shinji-II}, the authors worked in the framework of curvatons \cite{Curvaton}, and inflation was not necessary 
to produce the scale-invariant power spectrum. As a result,  the time of interest was assumed to be the period in which we have $H \gg M_*$. Thus,
 the quantization of their Lifshitz scalar $\phi$, which is responsible for generating the primordial curvature perturbations,
 can be carried out as  \cite{Shinji-II},
\bqn
\lb{2.50}
\zeta &\propto& \frac{\phi}{\mu},\nb\\
\phi\left(\mathbf{x},t\right)  &=& \frac{1}{(2\pi)^3}\int d^3k e^{-i\mathbf{k}\mathbf{x}}\left[u_k(t)\hat{a}_{\mathbf{k}} + u^*_k(t)\hat{a}^{\dagger}_{-\mathbf{k}}\right],\nb\\
u_k(t) &\propto& \frac{M_*}{k^{3/2}} \exp\left[-i \frac{k^3}{M^2_*}\int \frac{dt}{a^2}\right].
\eqn
The mode freezes after it leaves the sound horizon $(HM_*^3)^{-1/4}$, which is much smaller than the Hubble horizon in the inflation scenario, and gives rise to 
 a power spectrum,
\bqn
\lb{2.52}
P_{\zeta} =  \frac{M^2}{\left(2\pi\right)^2\mu^2}.
\eqn
This is quite different from the   expression 
\bq
\lb{2.52a}
P^{\text{GR}}_{\zeta} =  \frac{H^4}{\left(2\pi\right)^2\dot{\bar{\chi}}^2},
\eq
obtained with inflation in GR. In contrast, slow-roll inflation 
is required in our current model and the power spectrum was found to resemble the GR one \cite{Inflation}, given by
\bqn
\lb{2.55}
P_{\cal{R}} &\simeq&  {P^{\text{GR}}_{{\cal{R}}}} \Big[1- {\cal{O}}\left(\epsilon_{{\scriptscriptstyle \text{HL}}}\right) \Big], 
\eqn
where $\epsilon_{{\scriptscriptstyle \text{HL}}}\equiv {H^2}/{M^2_*} \ll 1$.

With the self-interaction at hand, and the relation $S|_{(3)}= - \int dt H|_{(3)}$, we perform the calculations of the bispectrum by using the in-in formalism 
\cite{Madalcena, Weinberg05},
\bqn
\lb{2.70a}
\left\langle \hat{Q}(t)\right\rangle &=&\Bigg\langle \Bigg[\bar{T} ~ \text{exp} \Big(i \int^t_{t_0} \hat{H}_{\text{I}}(t') dt'\Big)\Bigg]\nb\\
&& ~~~ \hat{Q}(t)\Bigg[T ~ \text{exp} \Big(-i \int^t_{t_0} \hat{H}_{\text{I}}(t') dt'\Big)\Bigg]\Bigg\rangle,\nb
\eqn
and find that to the  leading order in $H_{\text{I}}$ we have 
\bqn
\lb{2.70b}
&&\left\langle {\hat{\cal{R}}}_{\mathbf{k}_1}(t)\hat{{\cal{R}}}_{\mathbf{k}_2}(t)\hat{{\cal{R}}}_{\mathbf{k}_3}(t)\right\rangle \nb\\
& & ~~~  \simeq i \int^t dt'\left\langle  \left[\hat{H}|_{(3)}(t'),\hat{{\cal{R}}}_{\mathbf{k}_1}(t)\hat{{\cal{R}}}_{\mathbf{k}_2}(t)\hat{{\cal{R}}}_{\mathbf{k}_3}(t)\right]\right\rangle \nb\\
 & & ~~~ =  -i h^3 \left(2\pi\right)^3 \delta^3\left(\mathbf{k}_1 +\mathbf{k}_2 +\mathbf{k}_3\right) \nb\\
&&~~~ \Bigg\{6 V_3 \Big[k^2_1 k^2_2 k^2_3\Big]+ V_5 \Big[ \left(k^6_1 + k^6_2 + k^6_3\right) \nb\\ && ~~~~~~~~~~ - \left(k^4_1k^2_2 + k^2_1k^4_2+ \text{cyclic}\right)\Big]\Bigg\}\nb\\
&&~~~~~  \times \Big\{U(t';t) - \text{c.c.}\Big\} ,
\eqn
where
\bqn
\lb{2.71}
U(t';t) &\equiv& r^*_{k_1}(t)r^*_{k_2}(t)r^*_{k_3}(t)\nb\\  && \times \int^t \frac{dt'}{a^3(t')} r_{k_1}(t')r_{k_2}(t')r_{k_3}(t'). 
\eqn
$r_k(t)$ is the mode function for $\cal{R}$ \footnote{Quantization of the gauge invariant perturbation $\cal{R}$ was 
performed in \cite{Inflation} through the canonically normalized field $v=z{\cal{R}}, z^2 \equiv a^2h^2\beta_0$. 
For the simplicity of calculations, here we introduce the mode functions of $\cal{R}$, $r_k(\eta)$, which relates to $v_k(\eta)$ through the  relations,  
\bqn
\lb{2.80}
z(\eta) = \frac{v_k(\eta)}{r_k(\eta)} = \frac{v}{{\cal{R}}} .
\eqn
}. 
In writing down (\ref{2.70b}), we assumed that $d(V_{3,5})/dt=V'_{3,5} \dot{\bar{\chi}}=0$. We'll see that, once  this assumption is relaxed, it 
will generate more interesting features in the shapes of the bispectrum.

\section{Modified dispersion relations and the shapes of the bispectrum}
\renewcommand{\theequation}{4.\arabic{equation}} \setcounter{equation}{0}

To study the shapes of the bispectrum,  in this section we first consider the time evolution of the mode function in the (quasi-)de Sitter background.

\subsection{Time evolution of the mode function}

In the de Sitter background $a = -1/(H\eta)$, the equation of motion (EoM) of the mode function takes the form \cite{Inflation},
\bqn
\lb{3.10}
&& v''_k(\eta) + \Big(\omega^2 - \frac{2}{\eta^2}\Big) v_k(\eta) =0, \\
\lb{3.11}
&& \omega^2 = k^2\left(c^2_s + b_2 \frac{H^2}{M^2_*}k^2\eta^2+ b_3 \frac{H^4}{M^4_*}k^4\eta^4\right),
\eqn
where $c^2_s = b_1$ and $b_2, b_3$ are defined in \cite{Inflation}. Due to the time-dependence of the dispersion relation, the mode
function $v_{k}(\eta)$ (or equivalently $r_k(t)$) will not take a simple plane wave form as in  GR. Rather, its form will evolve with time. This 
complicates the calculations of the bispectrum considerably.

One may worry that, like in the case of \cite{Shinji-II}, relevant scales may have left the horizon at a time when the $k^6$ term dominated the dispersion relation. 
However, this seems not reasonable in the slow-roll inflation scenario. For the mode to leave the horizon at that time, two conditions have to be met,
\bqn
\lb{3.50a}
&& b_3\epsilon_{{\scriptscriptstyle \text{HL}}}^2 k^4\eta^4 \gg c^2_s \sim 1, \;\; \text{(UV regime)}   \nb\\
\lb{3.50b}
&& b_3\epsilon_{{\scriptscriptstyle \text{HL}}}^2k^6\eta^4 \ll \frac{a''}{a} = \frac{2}{\eta^2}, \text{(Super-horizon region)}. ~~~~~~
\eqn
This would indicate that $k^2 \eta^2 \ll 1$ and at the same time $b_3 \gg 1/(\epsilon_{{\scriptscriptstyle \text{HL}}}^2k^2 \eta^2) \gg 1$. 
However,  it was argued in \cite{Inflation} that $b_3$ in general is of order one. Therefore, we consider this case as physically not realistic.

\begin{figure}[t]
\centerline{\includegraphics[width=8cm]{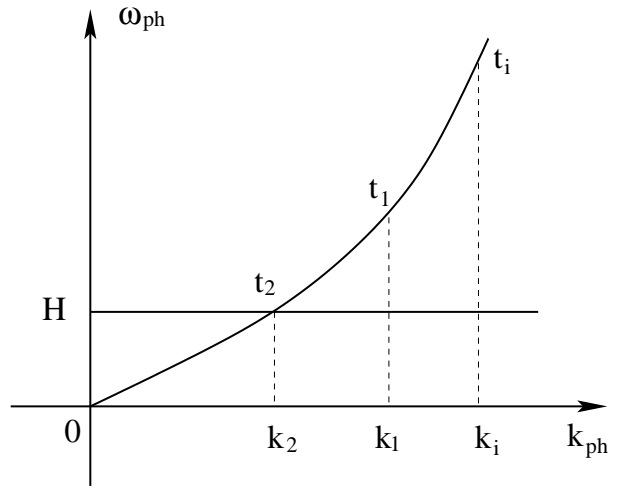}}
\caption{The evolution of $\omega_{\text{ph}} \equiv \omega_{k}/a$ vs $k_{ph} \equiv k/a$ in three different regions, where
Region I: $ t \in (t_i, t_1)$; 
Region II: $t \in (t_1, t_2)$; and Region III: $ t\in (t_2, t_0)$.}
\label{fig0}
\end{figure}

Then,  the evolution of the mode function during inflation has to be taken into account. To deal with this problem,  Brandenberger and Martin (BM)
proposed a matching procedure \cite{TransPl}. Three regions were identified for the evolution history of the mode function (See Fig. \ref{fig0}.
More divisions are possible 
given a specific model.). Region I is the region in which the UV effects dominate, alias $\omega \sim k^3$ in the present case. Then, the  solution of
 the EoM in this region   can be approximated with the Bessell functions of the first kind, 
\bq
\lb{3.70}
v^{\text{I}}_k(\eta) \simeq A_1 \sqrt{|\eta|}J_{\nu}\left[z(\eta)\right] + A_2 \sqrt{|\eta|}J_{-\nu}\left[z(\eta)\right].
\eq
In Region II,  the dispersion relation restores its relativistic form $\omega \sim k $, so that the mode function can be safely approximated with plane wave solutions,
\bq
\lb{3.80}
v^{\text{II}}_k(\eta) \simeq B_1 \exp\left[-i c_s k\eta\right] + B_2 \exp\left[i c_s k\eta\right].
\eq
Note that unlike in the case of the Bunch-Davies vacuum, since the mode function underwent a UV stage, both positive and negative frequencies appear in the 
mode function in this region. Region III is the super horizon region when the mode freezes. The initial conditions in Region I are the ones that minimize the energy
of the ground state of the field \cite{TransPl}, given, respectively, by Eqs.(\ref{D.10}) and (\ref{D.11}) in Appendix C of this paper,
 from which the   two constants $A_1$ and $A_2$ are fixed. The undetermined coefficients of the solutions in Regions II and III 
are fixed by matching conditions across each boundary of these regions, by requiring that  the mode function and its   first order time derivative  be  continuous.
The explicit expressions of these constants in terms of the initial conditions are given in Appendix C.

On the other hand,  the evaluation of the bispectrum all boils down to the following integration of the mode function,
\bqn
\lb{3.100}
r^*_{k_1}(t)r^*_{k_2}(t)r^*_{k_3}(t)\int^t \frac{dt'}{a^3(t')} r_{k_1}(t')r_{k_2}(t')r_{k_3}(t'). 
\eqn
A technical difficulty arises when the dispersion relation is of the form (\ref{3.11}) as no exact solutions exist.
Thus,  the matching procedure presented in \cite{TransPl} seems a natural choice in approximating the solution of the mode function to the EoM.
Below we evaluate the mode integration (\ref{2.71}) and calculate the bispectrum and its shapes more explicitly using the BM matching procedure.

\subsection{The Bispectrum}

Dividing the integration into three regions, as mentioned previously, one can see that only that over Regions I and II  need to be considered, as  Region III is the
 super horizon region, and the mode function gets  frozen out. Then, we find that
%
\bqn
\lb{3.150}
&& r^*_{k_1}(t)r^*_{k_2}(t)r^*_{k_3}(t)\int^t \frac{dt'}{a^3(t')}r_{k_1}(t')r_{k_2}(t')r_{k_3}(t') \nb\\
&& ~~~~ =  \Psi^*(t) \times \int^{t_1}_{t_i}\frac{dt'}{a^3(t')}  r^{\text{I}}_{k_1}(t')r^{\text{I}}_{k_2}(t')r^{\text{I}}_{k_3}(t')\nb\\
&& ~~~~~~~ + \Psi^*(t) \times \int^{t}_{t_1}\frac{dt'}{a^3(t')}\Psi(t') \nb\\
&&~~~  \equiv U^{\text{I}}\left(t'\in(t_i,t_1);t\right) + U^{\text{II}}\left(t'\in(t_1,t);t\right),
\eqn
where
\bqn
\lb{3.160}
\Psi(t) &=& r^{\text{II}}_{k_1}(t)r^{\text{II}}_{k_2}(t)r^{\text{II}}_{k_3}(t).
\eqn

To carry out the above integrations, we first note that in Region I,  $t \in (t_i, t_1)$, in contrast to the case considered in  \cite{Shinji-II} where the modes all take the 
form $\exp[-i\int dt' \omega/a^3] \sim \exp[-i k^3 \eta'^3] $, only those mode that are in the kernel of the integration take this asymptotic shape, 
while  $r^{\rm II}_{k_i}(t)$ takes the  form of plane waves. Therefore, in Region I we have, 
\bqn
\lb{3.170a}
&&U^{\text{I}}\left(t'\in(t_i,t_1);t\right) = \frac{M^2}{i} \sum_{\substack{
            x,y,z\\=1,2}} {\cal{B}}^*_{xyz} \sum_{\substack{
            l,m,n\\=1,2}}\frac{\sigma^{\text{I}}_{lmn}}{\mathscr{K}_{lmn}} ,\nb\\
 \lb{3.170b}
&& {\cal{B}}^*_{xyz} \equiv \tilde{B}^*_x (k_1)\tilde{B}^*_y (k_2)\tilde{B}^*_z(k_3)\exp\Big(-i c_s {\cal{K}}_{xyz} \eta\Big) \nb\\
&& \sigma^{\text{I}}_{lmn} \equiv  \tilde{A}_l(k_1)\tilde{A}_m(k_2)\tilde{A}_n(k_3)  \nb\\
							&&~~~~~~~~~~\exp \Bigg[\frac{i H^2}{3M^2}  \mathscr{K}_{lmn} \left(\eta^3_1-\eta^3_i\right) \Bigg],
\eqn
where  the coefficients $\tilde{A}$ and $\tilde{B}$ are given in Appendix C, and 
\bqn
\lb{3.171}
\mathscr{K}_{lmn} &\equiv& (-1)^l k^3_1+(-1)^m k^3_2+(-1)^n k^3_3, \nb\\
{\cal{K}}_{xyz} &\equiv& (-1)^x k_1 + (-1)^y k_2 + (-1)^z k_3.
\eqn

On the other hand, in Region II  we find that
\bqn
\lb{3.180a}
&& U^{\text{II}}\left(t'\in(t_1,t_2);t\right) = H^2 \sum_{\substack{
            x,y,z\\=1,2}} {\cal{B}}^*_{xyz}\sum_{\substack{
            l,m,n\\=1,2}}\sigma^{\text{II}}_{lmn}\nb\\
							&&~~~~~~~~~~~~~~~~~~~~ \times \Bigg\{\frac{\eta^2-\eta^2_1}{i {\cal{K}}_{lmn}} + \frac{2\left(\eta-\eta_1\right)}{{\cal{K}}^2_{lmn}} - \frac{2}{i 
							{\cal{K}}^3_{lmn}}\Bigg\} ,\nb\\
\lb{3.180b}
&& \sigma^{\text{II}}_{lmn} \equiv  \tilde{B}_l(k_1)\tilde{B}_m(k_2)\tilde{B}_n(k_3) 
\exp\Big[i c_s{\cal{K}}_{lmn} \left(\eta-\eta_1\right)\Big].\nb\\
\eqn
Putting all the above  together, we find that the bispectrum can be cast in the form, 
\bqn
\lb{3.185}
&& \left\langle {\cal{R}}_{\mathbf{k}_1}{\cal{R}}_{\mathbf{k}_2}{\cal{R}}_{\mathbf{k}_3}\right\rangle =   2 h^3 \left(2\pi\right)^3 \delta^3\left(\mathbf{k}_1 +\mathbf{k}_2 +\mathbf{k}_3\right) \nb\\
&& ~~~~~~  \times \Bigg\{6 V_3 \Big[k^2_1 k^2_2 k^2_3\Big]+ V_5 \Big[ \left(k^6_1 + k^6_2 + k^6_3\right) \nb\\ 
&& ~~~~~~~~~ - \left(k^4_1k^2_2 + k^2_1k^4_2+ \text{cyclic}\right)\Big]\Bigg\}
  \times\text{Im}\Big\{U^{\text{I}}+ U^{\text{II}}\Big\}.\nb\\
\eqn
Using the expression of the mode function $r_k(t)$ given  in Appendix C, we find that  the non-linearity parameter $f_{NL}$ can be estimated as,
\bqn
\lb{2.90a}
f_{\text{NL}} &=& \frac{\left\langle{\cal{R}}{\cal{R}}{\cal{R}}\right\rangle}{\left\langle{\cal{R}}{\cal{R}}\right\rangle^2}
\simeq \frac{\left(2\pi\right)^3}{\left[\left(2\pi\right)^3 2\pi^2\right]^2}\left(\frac{M_*}{h\sqrt{\beta_0}}\right)^6  \frac{k^6}{P^2_{\cal{R}}}\nb\\
&& \times \left[\frac{M_*^2}{k^3+k^3+k^3} + \frac{H^2\eta^2_1 \left(k+k+k\right)^3}{ \left(k+k+k\right)^3}\right]\nb\\
&& \times h^3 \left[6V_3\left(k^2k^2k^2\right) + V_5 \left(3k^6-6k^6\right)\right]\nb\\
&\simeq& {\cal{O}}\left(h^3 V_{3,5}\right)\left(\frac{M_*}{h}\right)^6\frac{10^{-5}M_*^2}{\left(4.9\times 10^{-5}\right)^4},
\eqn
where we have made the assumption $k_1=k_2=k_3=k$, $\beta_0, c_s \sim 1$ and used the fiducial COBE normalization \cite{COBE} of the power spectrum, from which
 we are able to reproduce the spectrum presented in GR under the assumption $H<<M_*$. 
Writing $M_*=10^{-n}M_{\text{pl}}$ and assigning $\epsilon \sim 10^{-2}$, when $n=2$, we obtain
\bq
\lb{2.90b}
f_{\text{NL}} \sim \frac{10}{\epsilon^3}  {\cal{O}}\left(h^3 V_{3,5}\right)\sim 10^{7} \times {\cal{O}}\left(h^3 V_{3,5}\right);
\eq
and  when $n=3$, we find that
 \bq
 \lb{2.90c}
 f_{\text{NL}} \sim \frac{10^{-7}}{\epsilon^3}  {\cal{O}}\left(h^3 V_{3,5}\right)\sim 10^{-1} ~ {\cal{O}}\left(h^3 V_{3,5}\right).
 \eq
Therefore, a large non-Gaussianity can be produced with a relative small $V_3$ and $V_5$, given that the new scale $M_{*}$ isn't much lower than $M_{\text{pl}}$.

\subsection{Shapes of the bispectrum}

Let us first note that, from (\ref{3.185}), the $k$-dependence of the spectrum, a.k.a the \emph{shape} and \emph{running}, depends not only on the action, but also on the form of the mode function. To study the $k$-dependence of the bispectrum in more detail, let us first look at the expressions of $U(t';t)$ in Eqs. (\ref{3.170a}) and (\ref{3.180a}). The corresponding conformal time $\eta$ is usually taken to be at very late of the inflation era, i.e.,  $k\eta \to 0$. To be simple, here we take $\eta=0$.

With the coefficients $\tilde{A}$ and $\tilde{B}$ given in Appendix C, the bispectrum is plotted 
 in Figures \ref{fig1} - \ref{fig4}   for various choices of the parameters. 
Figures \ref{fig1} and \ref{fig3} are for the choice of parameters (\ref{D.35a}), whereas Figures  \ref{fig2} and \ref{fig4}   are for the choice of parameters (\ref{D.35b}). Within each figure, four sub-cases are plotted.

From these figures, we can see that when $\delta_{\scriptscriptstyle{A}}$ and $\delta_{\scriptscriptstyle{B}}$ are taken to be zero, our result resembles the shape of $\alpha_2$ and $\alpha_3$ terms in \cite{Shinji-II}, despite the fact that our integration (\ref{3.150}) is actually different from theirs. This is quite understandable from the following considerations: when we make  $\delta_{\scriptscriptstyle{A}}=0=\delta_{\scriptscriptstyle{B}}$, only the positive (or negative, depending on the choice of sign for the initial condition) frequency modes exist in both regions. This makes the product of the six mode functions in (\ref{3.100}), which in general has $2^6$ terms since each of them has two branches, collapses into one single term. Substituting this into Eqs. (\ref{3.170a}) and (\ref{3.180a}), we obtain
\bqn
\lb{3.190}
U^{\text{I}}\left(t'\in(t_i,t_1);t\right)  &\propto& \frac{1}{\left(k_1k_2k_3\right)^3}\frac{1}{k^3_1+k^3_2+k^3_3}, \nb\\
U^{\text{II}}\left(t'\in(t_1,t);t\right) &\propto& \frac{1}{\left(k_1k_2k_3\right)^3}\frac{1}{\left(k_1+k_2+k_3\right)^3}.\nb\\
\eqn
We see that the contribution from Region I has the same $k$-dependence as the $\alpha_2$ and $\alpha_3$ terms in (4.5) and (4.6) of \cite{Shinji-II}. The contribution from Region II does not exist there, nor does it have the same $k$-dependence as in the relativistic cases.

The real difference comes in when we make either $\delta_{\scriptscriptstyle{A}}$ or $\delta_{\scriptscriptstyle{B}}$ nonzero. The impact of these on the shape of the bispectrum is the enhancement of the \emph{folded shape} \cite{Contractions} (or sometimes called the \emph{flattened shape} \cite{Holman}), namely, that the bispectrum peaks at the limit $k_2 + k_3 - k_1 \to 0$ and $k^3_2 + k^3_3 - k^3_1 \to 0$. A non-zero $\delta_{\scriptscriptstyle{A}}$ for the positive frequency choice of the initial condition results in the appearance of ``negative-frequency" modes in both $r^{\text{I}}_{k_1}(t)$ and $r^{\text{II}}_{k_1}(t)$, whereas a non-zero $\delta_{\scriptscriptstyle{B}}$ leads to a mixture only in $r^{\text{II}}_{k_1}(t)$. It's this mixture that makes the assumption which is usually kept in the standard choice of the 
Bunch-Davis (BD) vacuum \footnote{The initial conditions chosen in \cite{Shinji-II} are the BD-like vacuum.}, that is, only positive frequency modes appear in the mode function, invalid. In fact, when we take $\eta_i$ to be infinite past like in the BD vacuum, $\delta_{\scriptscriptstyle{A}}$ will be zero automatically as can be seen from its definition in (\ref{D.35c}). When we take $\delta_{\scriptscriptstyle{A}}=0$ but $\delta_{\scriptscriptstyle{B}}\neq0$, this is  similar to the case studied in \cite{Holman} and our result is consistent with theirs.

In addition, the choice of negative frequency $\tilde{A}_{-}$ essentially changes the sign of the bispectrum, and this is expected by comparing (\ref{D.35a}) with (\ref{D.35b})
\footnote{Note that our self-interaction terms (\ref{2.5}) have opposite sign w.r.t \cite{Shinji-II}. This explains why our shape reproduces theirs only when we take the negative frequency.}. 

We do not have an enhanced \emph{``squeezed" triangle signal} as in the minimum scenario (though the overall magnitude there is very small), nor do we have a large \emph{local form}, which is a typical result of multi-field models. This is because the interaction terms that could generate the local form are all suppressed by factors of either $c_1/M_{\text{pl}}$ or $\epsilon$ and are of sub-leading order [See discussions between (\ref{2.2b}) and (\ref{2.5})].

Now recall that in writing down (\ref{2.70b}), we have assumed that $V'_3$ and $V'_5$ are zero. However, this is not physically necessary since $V_3$ and $V_5$ appears neither in the background equations, nor in the linear perturbations  thus not constraint by the slow-roll conditions, and a strongly varying shape of $V_3$ and $V_5$ (not to be too strong to invalidate the perturbative expansion) would actually give both a higher non-linearity and new features in the bispectrum, such as the \emph{sinusoidal running} and \emph{resonant running}, as  shown in \cite{Contractions}.

Though the final integration was separated into two distinct periods, we would like to point out that this is a result of the matching procedure employed due to the lack of exact solutions for (\ref{3.10}) in our model. One should, in principle, evaluate the integration as a whole and study the shapes of the bispectrum. A possible solution is a development of the uniform approximation \cite{uniform} with some numerical integration techniques involved.
  \begin{figure}[hp!]
\centering
	\subfigure[$\delta_{\scriptscriptstyle{A}}=0=\delta_{\scriptscriptstyle{B}}$]
	{\label{fig:1a}\includegraphics[width=60mm]{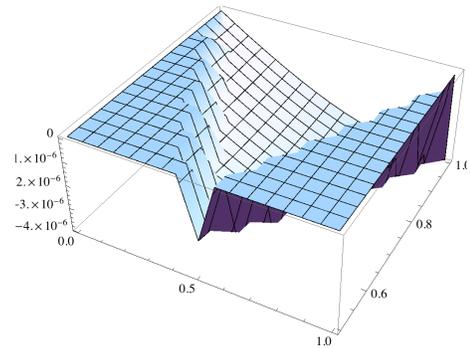}}\\
	\subfigure[$\delta_{\scriptscriptstyle{A}}=-0.1, ~\delta_{\scriptscriptstyle{B}}=0$]
	{\label{fig:1b}\includegraphics[width=60mm]{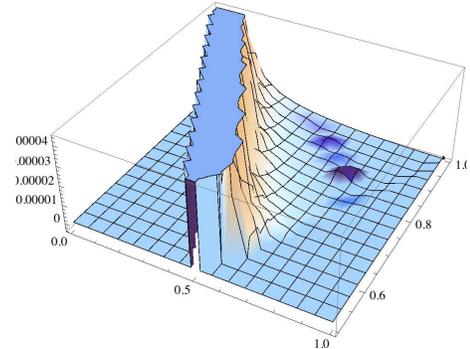}}\\
	\subfigure[$\delta_{\scriptscriptstyle{A}}=-0.1, ~\delta_{\scriptscriptstyle{B}}=0$]
	{\label{fig:1c}\includegraphics[width=60mm]{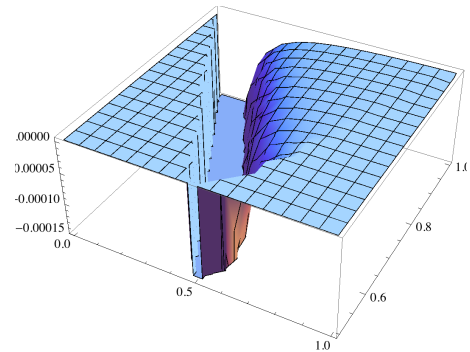}}\\
	\subfigure[$\delta_{\scriptscriptstyle{A}}=-0.1, ~\delta_{\scriptscriptstyle{B}}=-0.1$]
	{\label{fig:1d}\includegraphics[width=60mm]{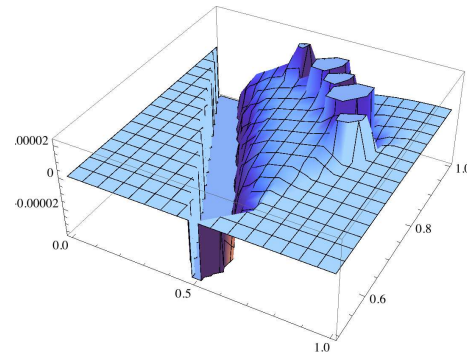}}
\caption{Shape of the bispectrum (truncated). $V_3$-term dominates, with the choice of the positive frequency.}
\lb{fig1}
\end{figure}

\begin{figure}[hp!]
\centering
	\subfigure[$\delta_{\scriptscriptstyle{A}}=0=\delta_{\scriptscriptstyle{B}}$]
	{\label{fig:2a}\includegraphics[width=60mm]{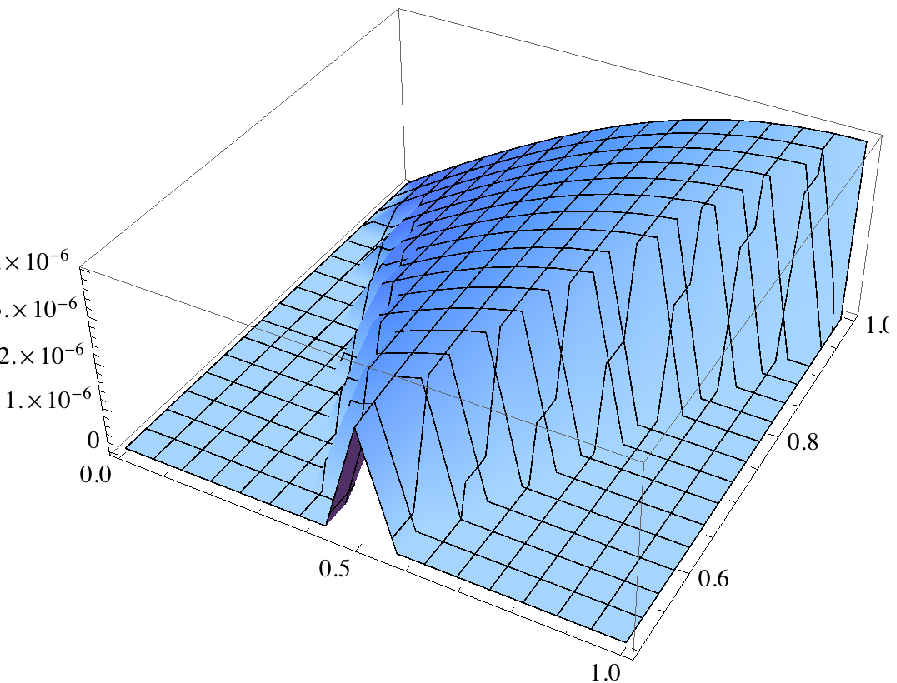}}\\
	\subfigure[$\delta_{\scriptscriptstyle{A}}=-0.1, ~\delta_{\scriptscriptstyle{B}}=0$]
	{\label{fig:2b}\includegraphics[width=60mm]{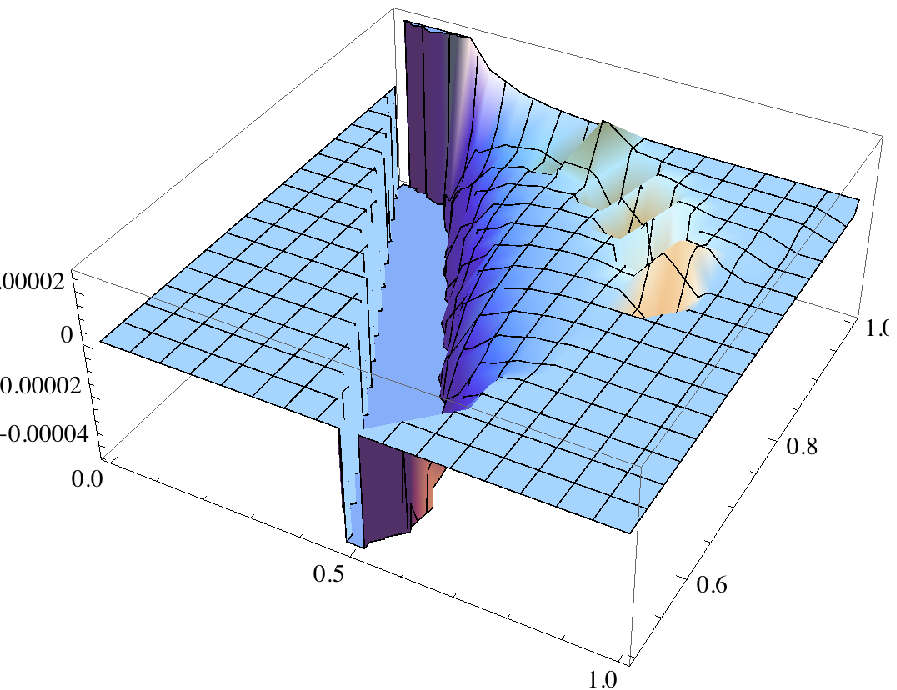}}\\
	\subfigure[$\delta_{\scriptscriptstyle{A}}=0, ~\delta_{\scriptscriptstyle{B}}=-0.1$]
	{\label{fig:2c}\includegraphics[width=60mm]{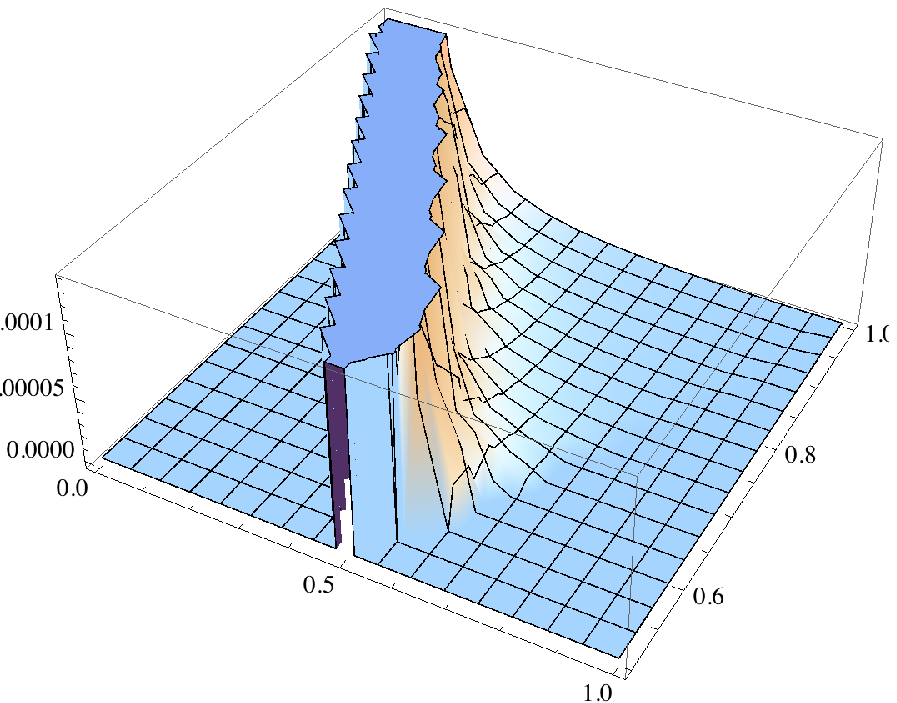}}\\
	\subfigure[$\delta_{\scriptscriptstyle{A}}=-0.1, ~\delta_{\scriptscriptstyle{B}}=-0.1$]
	{\label{fig:2d}\includegraphics[width=60mm]{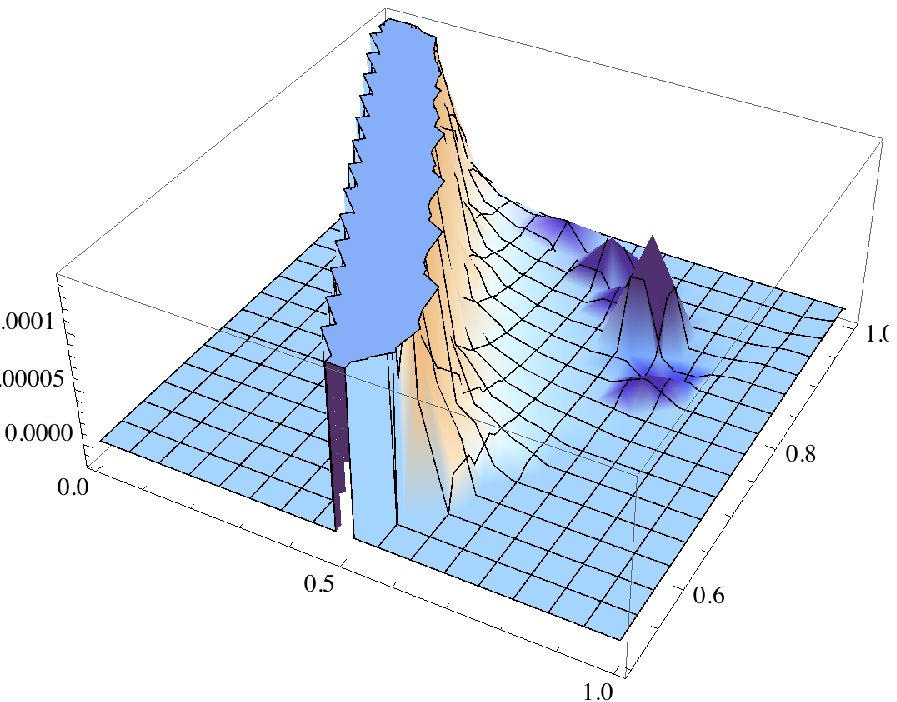}}
\caption{Shape of the bispectrum (truncated). $V_3$-term  dominates, with the choice of the   negative frequency.}
\lb{fig2}
\end{figure}

\begin{figure}[hp!]
\centering
	\subfigure[$\delta_{\scriptscriptstyle{A}}=0=\delta_{\scriptscriptstyle{B}}$]
	{\label{fig:3a}\includegraphics[width=60mm]{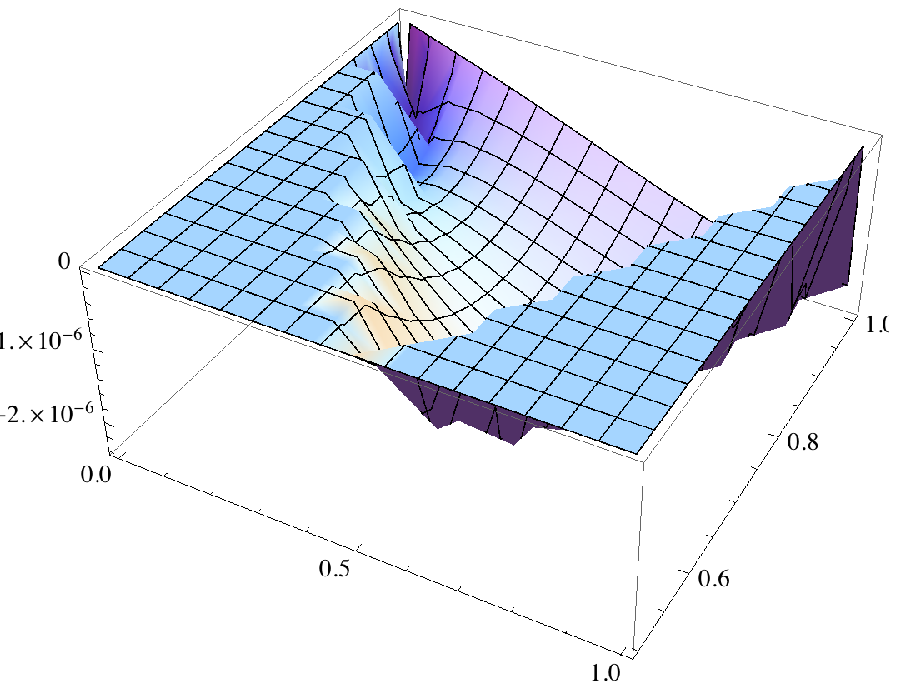}}\\
	\subfigure[$\delta_{\scriptscriptstyle{A}}=-0.1, ~\delta_{\scriptscriptstyle{B}}=0$]
	{\label{fig:3b}\includegraphics[width=60mm]{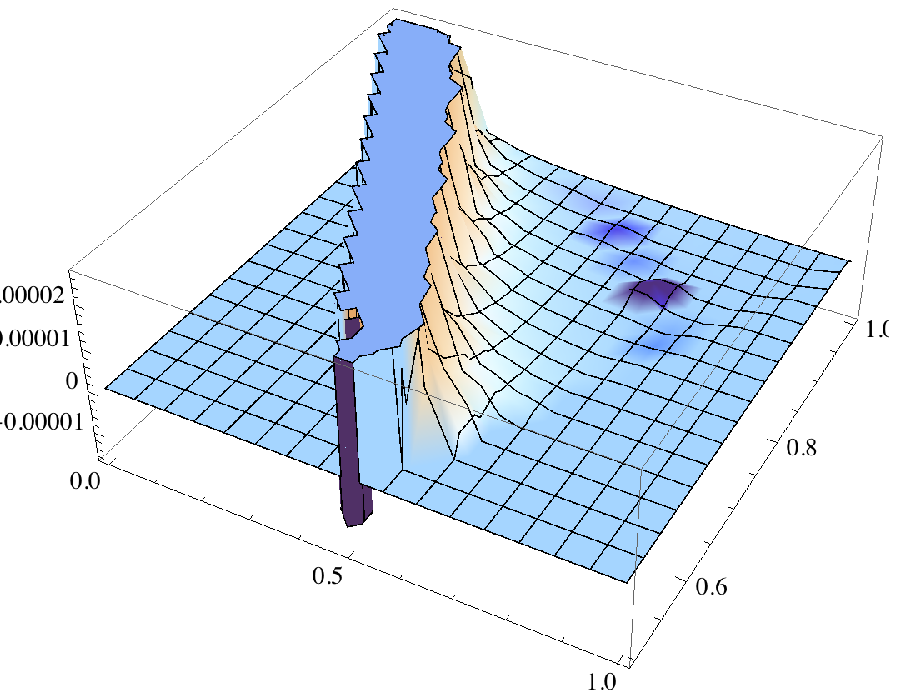}}\\
	\subfigure[$\delta_{\scriptscriptstyle{A}}=0, ~\delta_{\scriptscriptstyle{B}}=-0.1$]
	{\label{fig:3c}\includegraphics[width=60mm]{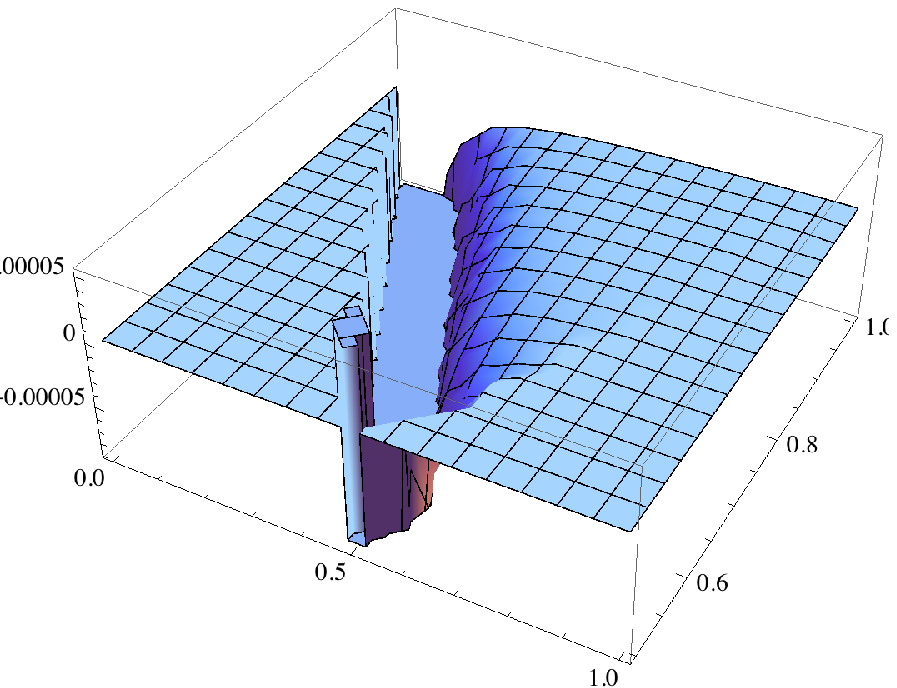}}\\
	\subfigure[$\delta_{\scriptscriptstyle{A}}=-0.1, ~\delta_{\scriptscriptstyle{B}}=-0.1$]
	{\label{fig:3d}\includegraphics[width=60mm]{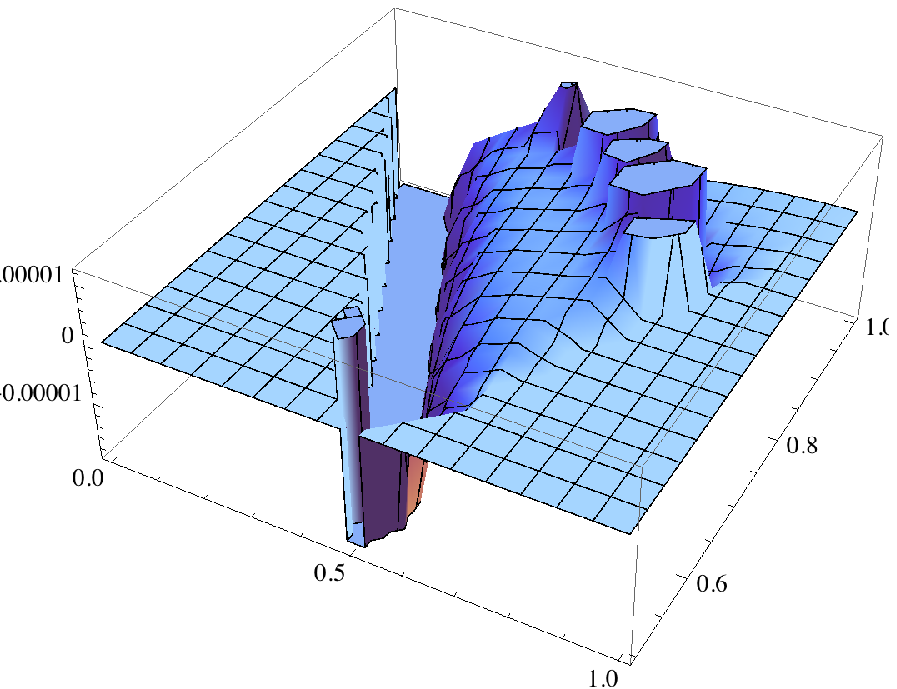}}
\caption{Shape of the bispectrum (truncated). $V_5$-term dominates, with the choice of the positive frequency.}
\lb{fig3}
\end{figure}

\begin{figure}[hp!]
\centering
	\subfigure[$\delta_{\scriptscriptstyle{A}}=0=\delta_{\scriptscriptstyle{B}}$]
	{\label{fig:4a}\includegraphics[width=60mm]{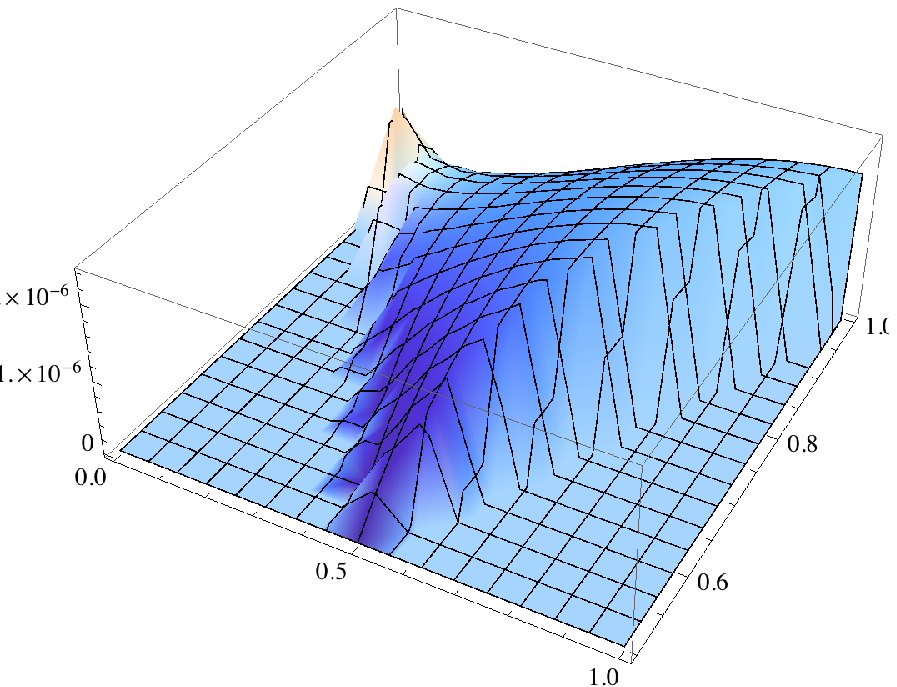}}\\
	\subfigure[$\delta_{\scriptscriptstyle{A}}=-0.1, ~\delta_{\scriptscriptstyle{B}}=0$]
	{\label{fig:4b}\includegraphics[width=60mm]{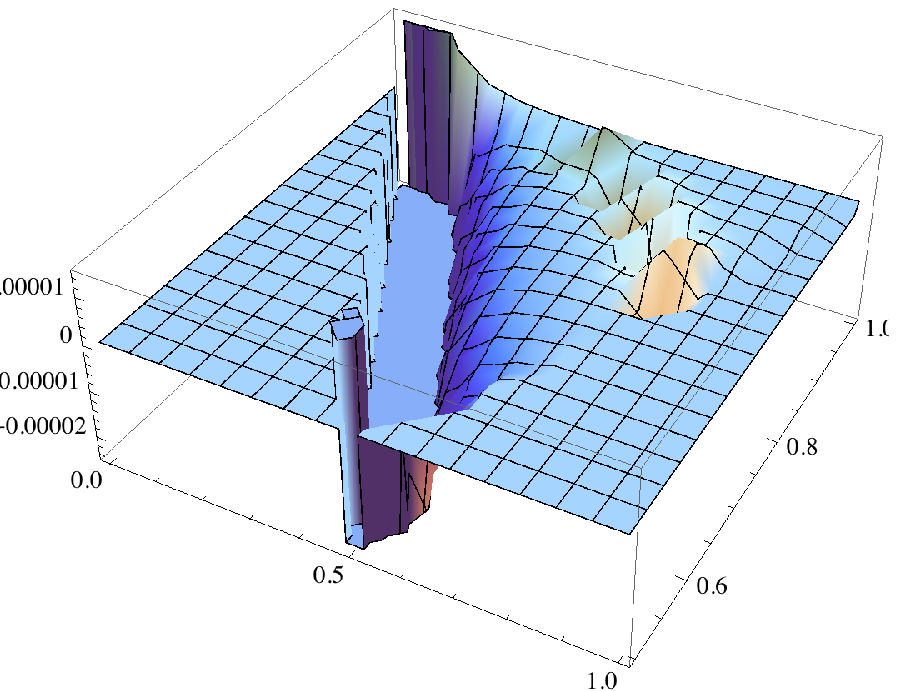}}\\
	\subfigure[$\delta_{\scriptscriptstyle{A}}=0, ~\delta_{\scriptscriptstyle{B}}=-0.1$]
	{\label{fig:4c}\includegraphics[width=60mm]{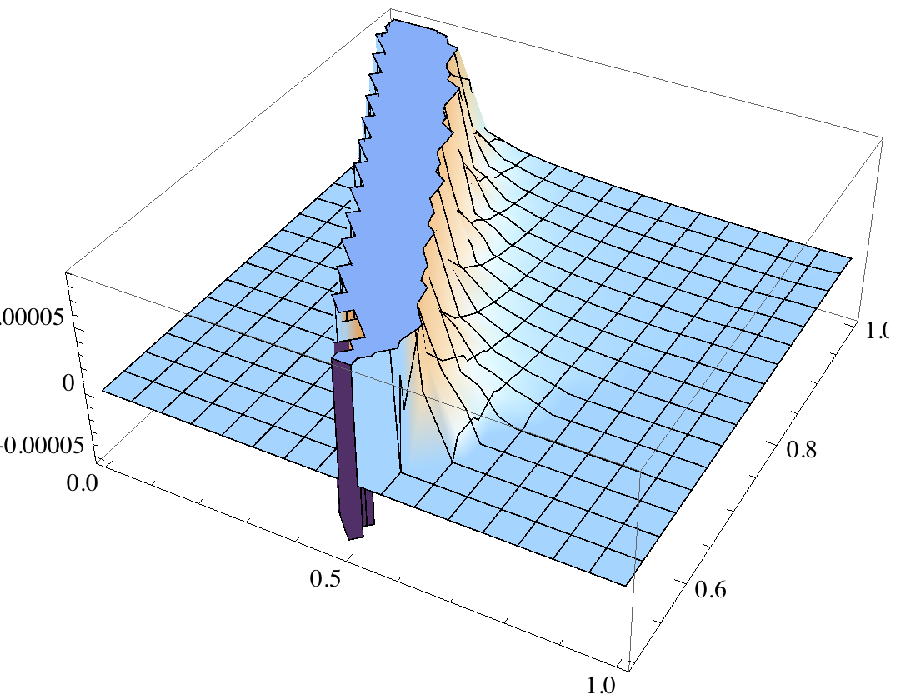}}\\
	\subfigure[$\delta_{\scriptscriptstyle{A}}=-0.1, ~\delta_{\scriptscriptstyle{B}}=-0.1$]
	{\label{fig:4d}\includegraphics[width=60mm]{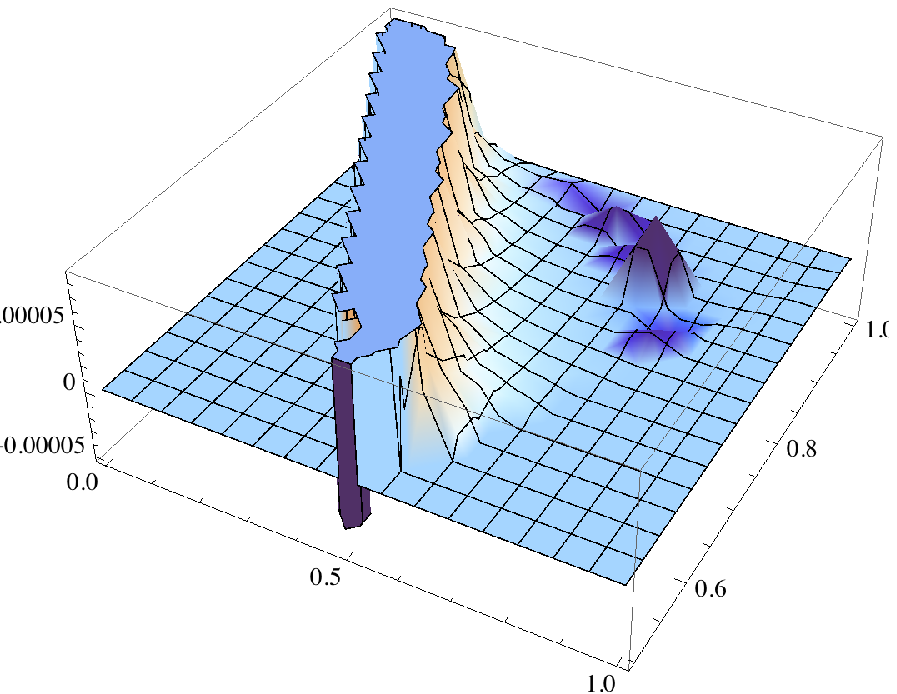}}
\caption{Shape of the bispectrum (truncated). $V_5$-term dominates, with the choice of the   negative frequency.}
\lb{fig4}
\end{figure}
\subsection{Projection onto factorizable templates}

To study the quantitative behavior of the shapes of the bispectrum, in general, factorable ansatz or template functions are utilized. Among them, three are of the most importance to us \cite{Contractions},
\bqn
\lb{3.200a}
T_{\text{Orth.}} \left(k_1,x,y\right)&\equiv& - 18\left(\frac{1}{x^3}+\frac{1}{y^3}+\frac{1}{x^3 y^3}\right) - \frac{48}{x^2y^2}\nb\\
&&  + 18\left(\frac{1}{x^2 y^3}+ \text{5 perms.}\right),\nb\\
T_{\text{Fold.}}\left(k_1,x,y\right)&\equiv& 6\left(\frac{1}{x^3}+\frac{1}{y^3}+\frac{1}{x^3 y^3}\right) + \frac{18}{x^2y^2}\nb\\
&&  - 6\left(\frac{1}{x^2 y^3}+ \text{5 perms.}\right),\nb\\
T_{\text{Equi.}}\left(k_1,x,y\right)&\equiv&-6\left(\frac{1}{x^3}+\frac{1}{y^3}+\frac{1}{x^3 y^3}\right) -\frac{12}{x^2y^2}\nb\\
&&  + 6\left(\frac{1}{x^2 y^3}+ \text{5 perms.}\right),
\eqn
where $x\equiv k_2/k_1, y\equiv k_3/k_1$. Then the bispectrum (\ref{3.185}) can be cast in the form, 
\bqn
\lb{3.203a}
&& \left\langle {\cal{R}}_{\mathbf{k}_1}{\cal{R}}_{\mathbf{k}_2}{\cal{R}}_{\mathbf{k}_3}\right\rangle
 \equiv \left(2\pi\right)^3 \delta^3\left(\mathbf{k}_1 +\mathbf{k}_2 +\mathbf{k}_3\right)\left(\frac{M_*}{h\sqrt{\beta_0}}\right)^6 \nb\\
&& ~~~~~~~~~~~~~~~~~~~~~~ \times T_{\text{HMT}}\left(k_1,k_2,k_3\right),
\eqn
where
\bq
\lb{3.203b}
T_{\text{HMT}} = c_{\text{O}}T_{\text{Orth.}}+c_{\text{E}}T_{\text{Fold.}}+c_{\text{F}}T_{\text{Equi.}}, 
\eq
with
\bqn
\lb{3.203c}
c_{\text{O}} &=&\frac{T_{\text{HMT}} \cdot T_{\text{Orth.}}}{T_{\text{Orth.}} \cdot T_{\text{Orth.}}}, ~~
c_{\text{F}} =\frac{T_{\text{HMT}} \cdot T_{\text{Fold.}}}{T_{\text{Fold.}} \cdot T_{\text{Fold.}}}, \nb\\
c_{\text{E}} &=&\frac{T_{\text{HMT}} \cdot T_{\text{Equi.}}}{T_{\text{Equi.}}\cdot T_{\text{Equi.}}},
\eqn
where the inner product is defined as \cite{Shinji-II}
\bqn
\lb{3.205}
T_1 \cdot T_2 &\equiv& \sum_{\mathbf{k}_i}\frac{T_1\left(k_1,k_2,k_3\right) T_2\left(k_1,k_2,k_3\right)}{P_{k_1}P_{k_2}P_{k_3}}\nb\\
&\propto& \int^1_0 x^4dx \int^1_{1-x} y^4dy T_1(1,x,y) T_2(1, x,y).\nb\\
\eqn

When $\delta_{\scriptscriptstyle{A}}=0=\delta_{\scriptscriptstyle{B}}$, the projection gives the following result
\bqn
\lb{3.210a}
c_{\text{O}} &=&  M^2 \left(1+\frac{H^2}{M_*^2}\right)\left(0.0033 V_3 - 0.0018 V_5\right)h^3,\nb\\
c_{\text{F}} &=&   M^2 \left(1+\frac{H^2}{M_*^2}\right)\left(0.0092 V_3 - 0.0004 V_5\right)h^3,\nb\\
c_{\text{E}} &=&   M^2 \left(1+\frac{H^2}{M_*^2}\right)\left(0.0063 V_3 - 0.0036V_5\right)h^3,\nb\\
\eqn
for the positive frequency,  and
\bqn
\lb{3.210b}
c_{\text{O}} &=& M^2 \left(1+\frac{H^2}{M_*^2}\right)\left(-0.0004 V_3 + 0.0002 V_5\right)h^3,\nb\\
c_{\text{F}} &=& M^2 \left(1+\frac{H^2}{M_*^2}\right)\left(-0.0001V_3 + 0.0000 V_5\right)h^3,\nb\\
c_{\text{E}} &=& M^2 \left(1+\frac{H^2}{M_*^2}\right)\left(-0.0008 V_3 + 0.0005V_5\right)h^3.\nb\\
\eqn
for the negative frequency. We see that the projection onto the equilateral template has the biggest magnitude for most of the cases. This is consistent with the observations in \cite{EFT} that derivative coupling favors the correlation between modes with similar momenta.

However, these analyses are only very rough approximations. 
This is because these three bases do not form a complete basis set, as pointed out in \cite{Shinji-II}, nor are they un-correlated, since we can compute correlations between them and get
\bqn
\lb{3.210c}
&& Corr\left(\text{Equi.}, \text{Orth.}\right) \sim 0.204,\nb\\
&& Corr\left(\text{Fold.}, \text{Orth.}\right) \sim -0.748,\nb\\
&& Corr\left(\text{Fold.}, \text{Equi.}\right) \sim 0.489,
\eqn
which indicate that there are strong correlations between the three, where
\bq
 Corr\left(\text{S1},\text{S2}\right) \equiv \frac{T_{\text{S1}}\cdot T_{\text{S2}}}{\sqrt{T_{\text{S1}}\cdot T_{\text{S1}}}\sqrt{T_{\text{S2}}\cdot T_{\text{S2}}}}.
 \eq

When $\delta_{\scriptscriptstyle{A}}$ and $\delta_{\scriptscriptstyle{B}}$ are not zero, the bispectrum, if taken naively, diverges at the folded limit $k_2 + k_3 - k_1 \to 0$ or $k^3_2 + k^3_3 - k^3_1 \to 0$, and the above inner product fails to converge. To understand the divergence, let us look at the integration (\ref{3.170a}) in  Region I more closely. The divergence occurs since,  in addition to (\ref{3.190}), we still have $(2^6-2)$-terms of the form, 
\bqn
\lb{3.215a}
\left(\delta_{\scriptscriptstyle{A}}\right)^m\left(\delta_{\scriptscriptstyle{B}}\right)^n\frac{1}{k^3_x + k^3_y - k^3_z}.
\eqn
The denominator appears through the $d\eta'$ integration in (\ref{3.150}). When the denominator approaches zero in the folded limit, if $\delta_{\scriptscriptstyle{A}}$ and $\delta_{\scriptscriptstyle{B}}$ are not exactly zero, these terms blow up. The divergences must be regulated away if we want to have a quantitative estimate of the projection. A possible solution is to introduce a cut-off so that the integration yields,
\bqn
\lb{3.215b}
\left(\delta_{\scriptscriptstyle{A}}\right)^m\left(\delta_{\scriptscriptstyle{B}}\right)^n\frac{1}{k^3_2 + k^3_3 - k^3_1 + k^3_c}
\eqn
where 
\bqn
\lb{3.215c}
k_c= k_c\left(\delta_{\scriptscriptstyle{A}},\delta_{\scriptscriptstyle{B}}\right),
\eqn
in order to keep (\ref{3.215b}) finite. As noted in \cite{Contractions}, this regularization is highly model-dependent and requires some systematic study. Putting this divergence aside, we would expect that the projection onto the folded base would be enhanced significantly, in comparing with (\ref{3.210a}) and (\ref{3.210b}).

\section{Conclusions and Remarks}
\renewcommand{\theequation}{5.\arabic{equation}} \setcounter{equation}{0}
We have studied in this paper non-Gaussianity of the primordial perturbations in a single scalar  field with the slow-roll 
conditions  in the framework of the Ho\v{r}ava-Lifshitz gravity with the projectability condition and an arbitrary coupling 
constant $\lambda$, refferred to as the HMT model.

With some reasonable assumptions, we have found that the leading order terms in self-interaction of the HMT
model are of order $\epsilon^{3/2} \hat\alpha_{n}$, where $\hat\alpha_{n}$ are dimensionless constants, defined as
$V_n \equiv \hat{\alpha}_{n}/ M^4_{\text{pl}}$, and $V_n$ are the coupling coefficients  of  sixth order derivative operators
of the scalar field, and have no contributions to the power spectra and indices, as shown explicitly  in \cite{Inflation}. 
Clearly, by properly choosing those coefficients, a large non-Gaussianity is possible. 
This is different from the standard result of minimum scenario in general relativity, where the interaction terms are  of order $\epsilon^2$
\cite{Madalcena,Weinberg05}.

We have also investigated how the modified dispersion relation affects the evolution of the mode function and in turn the bispectrum, 
using the matching method proposed in \cite{TransPl}. 
By dividing the history of inflation into three regions, in which the dispersion relation  takes different asymptotic forms,
 the gauge-invariant scalar perturbation  ${\cal{R}}$ has  different asymptotic solutions. 
In particular, we have found that the mode function is in general a superposition of oscillatory functions. 
This is different from the standard  choice of the Bunch-Davis vacuum, 
where only one positive frequency branch of the plane wave is selected, and  results in an enhancement of the 
folded shape in the bispectrum.

Due to the existence of a UV region, the bispectrum is enhanced, and gives rise to   a large nonlinearity parameter 
$f_{\text{NL}}$ [cf. Eq.(\ref{2.90a})],  as long as    $M_*$, above which the non-linear terms in the dispersion relation dominate, isn't much 
lower than the Planck scale [See (\ref{2.90b}) and (\ref{2.90c})].

We would also like to summarize the assumptions made along the analysis, as the invalidation of any of these assumptions will certainly change some of the conclusions.
In particular, in obtaining the leading order term in the self-interaction, we made the assumption that $c_1 \sim M_* \ll M_{\text{pl}}$. This eliminates a great number of terms which are not of order $\delta\chi^3$ in the third order expansion, and makes all terms that will give non-local effect (or local shape bispectrum) of sub-leading order. However, the phenomenological upper bound on $M_*$ for the projectable version of the HL theory is not established yet \cite{LMW}. Dropping this assumption will bring back a lot of terms and enhance the local shape signal in the bispectrum.

We've also kept the assumption made in \cite{Inflation} that $V'_n \simeq 0$. For $V_1, V_2$ and $V_4$, this assumption further eliminate some possible contributions to the leading order effect. For $V_3$ and $V_5$, as noted in the discussion at the end of Part.C in Section IV, if this assumption is dropped, new features  (sharp  and periodic shapes) will appear. 
The condition $H/M_*<<1$ has been also assumed to obtain an expression of the power spectrum similar to that of GR. However, if this is dropped out, the power spectrum will receive large corrections, as noted in \cite{Inflation}.

By dividing the time of interest into three regions, we have assumed that the period of dominance of the $k^4$ term in the dispersion relation  is so short that its effect can be incorporated into the small parameter $\delta_{\scriptscriptstyle{B}}$. This is a strong assumption, as noted in \cite{NaD, PPGW}.  During this period of time, $\omega^2_k$ may actually go below $H^2$ and then the mode function will be no longer  oscillating, but grow with the scale factor $a$. 
This is an important problem and deserves further investigations.

\section*{Acknowlodgements}
We would also like to express our gratitude to V. H. Satheeshkumar, who checked carefully some of our calculations, and 
to Shinji Mukohyama for valuable discussions and suggestions. 
AW and YH would like to thank Zhejiang University of Technology for the hospitality where most of the work was done.
AW is supported in part by the DOE  Grant, DE-FG02-10ER41692.

\section*{Appendix A:  Scalar perturbations in HMT Model}
 \renewcommand{\theequation}{A.\arabic{equation}} \setcounter{equation}{0}
For the reader's convenience, we present here the  linearized perturbations and the corresponding constraints and field equations that were first obtained in \cite{Inflation}. 
These will be used in obtaining the cubic action of  self-interaction. Below we shall choose the Newtonian quasi-longitudinal gauge (\ref{gauge})
\footnote{Note that the two popular gauges in GR:\\ i)$E=0=\delta\chi, B\neq0, \phi\neq0$ and $\zeta\neq0$;\\ ii)$E=0=\zeta, B\neq0, \phi\neq0$ and 
$\delta\chi\neq0$\\ are not possible in our model because of (\ref{b8}).}.
The variable $\psi$ defined as part of the perturbations to the 3-metric $g_{ij}$ in that paper is related to $\zeta$ used here via the relation, 
\bq
\lb{b1}
 \psi\simeq -\zeta.
\eq
The constraint equations to first order are then given by, 
\bqn
\lb{b5}
&&\int d^{3}x\Bigg\{\partial^2\psi
- \frac{1}{2}\big(3\lambda -1\big){\cal H}
\Big[3\psi' +\partial^2B\Big]\Bigg\} \nb\\ 
&& =4\pi G\int d^{3}x\Bigg\{f\bar{\chi}' \delta{\chi}' + \Big(a^2{V}' + \frac{V_{4}}{a^{2}} \partial^4\Big) \delta \chi\Bigg\}, ~~~~~\\
 \lb{b6}
 && (3\lambda -1){\psi}'    - (1- \lambda)\partial^2 B = 8\pi G f \bar{\chi}' \delta \chi, \\
  \lb{b7}
& & 2{\cal{H}}\psi +  \big(1-\lambda\big)\big(3\psi'   + \partial^2B\big) 
\nb\\
&&   ~~~~~ = 8\pi G\Big[\big({c}'_{1} \bar{\chi}'+ {c}_{1}{\cal H} - f \bar{\chi}'\big) \delta \chi + {c}_1  \delta \chi'\Big] ,\\
 \lb{b8}
&& \psi = 4\pi G {c}_1  \delta \chi.
\eqn
The trace- and traceless-dynamical equations are, respectively, given by
\bqn
\lb{b10}
 &&  \psi'' + 2{\cal{H}}\psi'   + \frac{1}{3}\partial^{2}\big(B' + 2{\cal{H}}B\big)\nb\\ 
& & ~~~~
 - \frac{2}{3(3\lambda-1)}\Bigg(1 + \frac{\alpha_{1}}{a^{2}}\partial^{2}
   + \frac{\alpha_{2}}{a^{4}}\partial^{4}\Bigg)\partial^{2}\psi \nb\\
   & & ~~~~ + \frac{2}{3(3\lambda - 1)a}\partial^{2}\big(\hat{A}\psi - \hat{\delta{A}} \big)\nb\\ 
   & & ~~~~  = \frac{8\pi G }{3\lambda-1}\Big(f\bar{\chi}' \delta \chi' - a^2{V}'  \delta \chi\Big), \\
\lb{b11}
 &&  \psi - \big(B' + 2{\cal{H}}B\big)  
+ \frac{1}{a^{2}}\Big(\alpha_{1} + \frac{\alpha_{2}}{a^{2}}\partial^{2}\Big)\partial^2\psi\nb\\
& & ~~~ -\frac{1}{a}\Big(\hat{A}\psi  - \hat{\delta{A}} \Big) = 0,
\eqn
where
 \bqn
 \lb{b12} 
 \alpha_{1} &\equiv& \frac{8g_{2} + 3g_{3}}{M^{2}_{\text{pl}}/2},\;\;\;
   \alpha_{2} \equiv  \frac{8g_{7}-3g_{8}}{M^{4}_{\text{pl}}/4}.
 \eqn
 The modified Klein-Gordon equation reads
 \bqn
 \lb{b13}
&& f \Big\{\delta \chi'' + 2 {\cal H} \delta \chi' - \bar{\chi}' \left[3 \psi' +\partial^2 B\right]\Big\} + a^2V''\delta\chi\nb\\
 &&   ~~~~ = 2 \Bigg(\frac{1}{2}+{V}_{1} -\frac{{V}_{2}+{V}'_{4}}{a^2}\partial^2 -\frac{{V}_{6}}{a^4} \partial^4\Bigg)\partial^2 \delta \chi\nb\\
 && ~~~~~~~~~~ +\frac{1}{a} \partial^2\Big[2\hat{A}\left({c}'_{1} - {c}_{2}\right)  \delta \chi  + c_1 \hat{\delta{A}}\Big].
 \eqn
Solving the above constraints and field equations yields $B$ and other variables in terms of $\zeta$,
\bqn
\lb{b20.a}
B &=& \frac{1}{|c^2_\psi|}\left(\partial^{-2}\zeta'\right) - \frac{2f\bar{\chi}'}{(1-\lambda)c_1}\left(\partial^{-2}\zeta\right), \\
\lb{b20.b}
\frac{\delta \hat{A}}{a} &=& \left[1 + \frac{\alpha_1}{z^2}\partial^2 + \frac{\alpha_2}{a^4}\partial^4-\bar{A}\right]\zeta +  \frac{1}{|c^2_\psi|}\left(\partial^{-2}\zeta''\right) \nb\\
&&  + \left[\frac{2f\bar{\chi}'^2}{(1-\lambda)c_1}\frac{c'_1}{c_1} + \frac{2V'}{a^2(1-\lambda)c_1}\right] \left(\partial^{-2}\zeta\right)\nb\\
&&  - \left[\frac{2f\bar{\chi}'}{(1-\lambda)c_1} -\frac{2{\cal{H}}}{|c^2_\psi|}\right]\left(\partial^{-2}\zeta'\right), \\
\lb{b20.c}
\delta \chi &=& -\left(4\pi G c_1\right)^{-1}\zeta,
\eqn
where $c^2_\psi \equiv (\lambda-1)/(1-3\lambda)$, and $\left(\partial^{-2}\partial^2\right) \zeta = \zeta$.

The quantity 
\bqn
\lb{b30}
{\cal{R}} &\equiv& \psi + \frac{H}{\dot{\bar{\chi}}}\delta\chi,\\
&=& -\Big(1+\frac{H}{4\pi G c_1 \dot{\bar{\chi}}}\Big)\zeta,
\eqn
often referred to as the comoving curvature perturbation, is also gauge-invariant. It can be shown that in terms of this quantity, the free action can be written as 
\bqn
\lb{b31}
&& S^{(2)} = \frac{1}{2}\int d\eta d^3 x a^2 h^2\Big[\beta_0 {\cal{R}}'^2 -\beta_4 {\cal{R}}^2 -  \beta_1 (\partial_i{\cal{R}})^2\nb\\
&& ~~~~~~~~~~~~~~~~~~~~~~~~~~~~~~~~ -  \beta_2 (\partial^2 {\cal{R}})^2 - \beta_3 (\partial_i\partial^2 {\cal{R}})^2 \Big], \nb\\
\eqn
where 
\bqn
\lb{b32}
\beta_0 &=& f +4 \pi G c^2_1 / |c^2_{\psi}|,\nb\\
\beta_1 &\equiv&  1+ 2 V_1 + 2 \bar{A} (c_1' - c_2) - 4\pi G c^2_1 (1-\bar{A}), \nb\\
\beta_2  &\equiv& \frac{2}{a^2} \left(V_2 + V_4' + 2\pi G c^2_1 \alpha_1\right), \nb\\
\beta_3 &\equiv& - \frac{2}{a^4} \left(V_6 + 2\pi G c^2_1 \alpha_2\right), \nb\\
\beta_4 &\equiv& \beta_0 {\cal{Q}} - \beta_0 \frac{h'^2}{h^2} + \frac{\left(a^2\beta_0hh'\right)'}{a^2 h^2},\nb\\
{\cal{Q}} &\equiv& \frac{1}{\beta_0}\Bigg\{a^2 V'' + \frac{4 \pi G c_1 c''_1 \bar{\chi}'^2 }{|c_{\psi}^{2}|} -\frac{8 \pi G}{\lambda -1}f\bar{\chi}'^2\Big(f- {c_1}'\Big)\nb\\
&& -4 \pi G c_1 a^2 V' \left(3 +\frac{ c'_1}{f |c_{\psi}^{2}|} - \frac{1}{|c_{\psi}^{2}|}\right)\Bigg\},\nb\\
h &\equiv& \left(4\pi G c_1 + \frac{H}{\dot{\bar{\chi}}}\right)^{-1} = \frac{\delta \chi}{{\cal{R}}}.
\eqn
After introducing the normalized variable
\bq
\lb{b33}
v \equiv  z {\cal{R}}, \;\;\;
z^2 \equiv a^2 h^2 \beta_0,
\eq
the classical equation of motion for the mode function $v_k$ takes the form, 
\bq
\lb{b34}
v''_k + \left(\omega^2_k  + m^2_{\text{eff}}\right) v_k =0, 
\eq
where
\bqn
\lb{b35}
\omega^2_k  &=& \frac{k^2}{\beta_0} \left(\beta_1 + {\beta_2} k^2 + {\beta_3} k^4\right), \\
- m^2_{\text{eff}} &\equiv& \frac{z''}{z} - \frac{\beta_4}{\beta_0}.
\eqn
\section*{Appendix B:  Self interaction of the inflaton}
 \renewcommand{\theequation}{B.\arabic{equation}} \setcounter{equation}{0}
Under the Newtonian quasi-longitudinal gauge (\ref{gauge}), with perturbations given by (\ref{2.0a}),
we find that the action can be written in the form, 
\bqn
\lb{C2}
S|_{(3)} &=& S_g|^{\text{GR}}_{(3)} + S_\chi|^{\text{GR}}_{(3)} \nb\\
&&  + S_g|^{\text{HC}}_{(3)} + S_A|_{(3)} + S_\chi|^{\text{HL}}_{(3)}, 
\eqn
where the ``GR parts" are given by \footnote{Note that though this is labeled as the ``GR part", it cannot reproduce the exact expression of GR, as a result of the difference in symmetry.},
\bqn
\lb{C3.1}
S_g|^{\text{GR}}_{(3)} &=& \frac{1-3\lambda}{16\pi G} \int d\eta dx a^2 \Bigg\{\frac{27}{2} {\cal{H}}^2 \zeta^3 \nb\\
&& ~~~~ + 9 {\cal{H}} \zeta^2 \left(3 \zeta' - \partial^2B\right) - 2 \zeta' \left(\partial_k B \partial^k \zeta\right) \nb\\
&& ~~~~ + 3 \Big[3\zeta\left(\zeta'\right)^2 - 2{\cal{H}}\zeta\left(\partial_kB\partial^k\zeta\right) - 2 \zeta\zeta'\partial^2B\Big]  \nb\\
&& ~~~~ + \frac{3}{1-3\lambda}\zeta \left(\partial_{ij}B\right)^2 -\frac{3\lambda}{1-3\lambda} \zeta \left(\partial^2 B\right)^2  \nb\\
&& ~~~~  \nb\\
&& ~~~~ -\frac{1+ \lambda}{1-3\lambda} 2 \left(\partial^2 B\right) \left(\partial_k B \partial^k \zeta\right)  \nb\\
&& ~~~~ + \frac{8}{1-3\lambda} \left(\partial_{ij} B \partial^i B \partial^j \zeta\right)\Bigg\}\nb\\
&&  - \frac{1}{16 \pi G} \int d\eta dx a^2 \Big\{9\Lambda a^2 \zeta^3 + \zeta^2 \partial^2 \zeta\Big\},
\eqn
\bqn
\lb{C3.2}
S_{\chi}|^{\text{GR}}_{(3)} &=& \int d\eta dx a^2 \Bigg\{\frac{9\zeta^3}{2}\left[ \frac{f(\lambda)}{2}\bar{\chi}'^2 - a^2V\right] \nb\\
&& ~~~~~~ + \frac{9\zeta^2}{2} \Big[f(\lambda)\bar{\chi}'\delta\chi' - a^2 V'\delta\chi \Big]\nb\\
&& ~~~~~~ + \frac{3\zeta}{2}\Big[ f(\lambda)\left(\delta\chi'\right)^2 - a^2 V''\delta\chi^2\Big] \nb\\
&& ~~~~~~ -3 \zeta f(\lambda)\bar{\chi}' \left(\partial_k B\right)\left(\partial^k\delta\chi\right)\nb\\
&& ~~~~~~ - f(\lambda)\delta\chi' \left(\partial_k B\right)\left(\partial^k\delta\chi\right) \nb\\
&& ~~~~~~ - a^2 \frac{V'''}{6}\left(\delta\chi\right)^3\Bigg\}. 
\eqn
And the ``HL gravitational  part" is given by, 
\bqn
\lb{C3.3}
S_g|^{\text{HC}}_{(3)} &=& \frac{1}{16\pi G} \int d\eta dx a^2 \Bigg\{-a^2 \mathscr{L}_V|^{\text{HC}}_{(3)} \nb\\
&& ~~~~ - \left(3\zeta\right) \frac{2}{a^2 M^2_{\text{pl}}} \Big[\left(16 g_2 + 5 g_3\right)\left(\partial^2\zeta\right)^2 \nb\\
&& ~~~~~~~~~~~~~~~ + g_3 \left(\partial_{ij} \zeta\right)^2\Big] \nb\\
&& ~~~~- \left(3\zeta\right)\Big( \frac{2}{a^2M^2_{\text{pl}}}\Big)^2 \Big[16 g_7 \left(\partial^2\zeta\right)\left(\partial^4\zeta\right) \nb\\
&& ~~~~~~~ + 5 g_8 \left(\partial_k \partial^2 \zeta\right)^2 + g_8 \left(\partial_{ijk} \zeta\right)^2\Big]\Bigg\},\nb\\
\eqn
where
\bqn
\lb{C3.4}
a^2 \mathscr{L}_V|^{\text{HC}}_{(3)} &=& 
 \frac{2}{a^2 M^2_{\text{pl}}}\Big[\left(-64 g_2 - 20 g_3\right) \zeta \left(\partial^2 \zeta\right)^2  \nb\\
&& ~~~~ + \left(16 g_2 + 6 g_3\right) \left(\partial^2 \zeta\right)\left(\partial_k \zeta\right)^2  \nb\\
&& ~~~~ - 2 g_3 \left(\partial_i\zeta \partial_j\zeta\partial^{ij}\zeta\right) - 4 g_3 \zeta \left(\partial_{ij}\zeta\right)^2\Big] \nb\\
&&+ \Big( \frac{2}{a^2M^2_{\text{pl}}}\Big)^2 \Big[ -12 g_8 \left(\partial_{ij}\zeta \partial_k\zeta \partial^{ijk}\zeta\right) \nb\\
&& ~~~~ + 16 g_8 \left(\partial_i\zeta \partial_j\partial^2\zeta \partial^{ij}\zeta\right) \nb\\
&& ~~~~ - \left(32 g_7 + 20 g_8\right) \left(\partial^2\zeta \partial_k\zeta \partial^k\partial^2\zeta\right) \nb\\
&& ~~~~ - 30 g_8 \zeta \left(\partial_k\partial^2\zeta\right)^2 - 6 g_8 \zeta \left(\partial_{ijk}\zeta\right)^2 \nb\\
&& ~~~~ - \left(64 g_4 + 20 g_5 + 6 g_6 + 32 g_7\right) \left(\partial^2\zeta\right)^3 \nb\\
&& ~~~~ + \left(16 g_7 - 4 g_5 - 3 g_6\right) \left(\partial^2\zeta\right) \left(\partial_{ij}\zeta\right)^2 \nb\\
&& ~~~~ - g_6 \left(\partial^{ij}\zeta \partial_{jk}\zeta \partial_i\partial^k\zeta\right) \nb\\
&& ~~~~ - 96 g_7 \zeta \partial^2\zeta \partial^4\zeta + 8 g_7 \partial^4 \zeta \left(\partial_k \zeta\right)^2\Big].\nb\\
\eqn
The ``gauge part" is given by, 
\bqn
\lb{C3.5}
S_A|_{(3)} &=& \frac{1}{16\pi G} \int d\eta dx a^2 \Bigg\{2 \bar{A} \zeta^2 \left(\partial^2\zeta\right) +2 \bar{A} \zeta  \left(\partial_k \zeta\right)^2  \nb\\
&& ~~~~ + 4 \zeta \left(\partial^2\zeta\right) \delta A +2 \left(\partial_k \zeta\right)^2 \delta A \Bigg\}.
\eqn
Finally, the ``HL matter part" is given by
\bqn
\lb{C3.6}
S_{\chi}|^{\text{HL}}_{(3)} &=& \int d\eta dx a^2 \Bigg\{\frac{9\zeta^2}{2} \Big[- \frac{V_4}{a^2}\left(\partial^4\delta\chi\right) + \bar{A}c_1 \left(\partial^2\delta\chi\right)\Big]\nb\\
&& ~~~~~~ + \left(3a^2\zeta\right)\Big[\mathscr{L}_{\chi}|^{\text{HL}}_{(2)} \Big] + a^2 \Big[\mathscr{L}_{\chi}|^{\text{HL}}_{(3)} \Big]\Bigg\},\nb\\
\eqn
with 
\bqn
\lb{C3.7}
\mathscr{L}_{\chi}|^{\text{HL}}_{(2)} &=& - V_4 \Big[{\cal{P}}_2|_{(2)}\Big] - \left(\frac{1}{2} + V_1\right) a^{-2}\left(\partial_k\delta\chi\right)^2 \nb\\
&& - \frac{V_2}{a^4}\left(\partial^2\delta\chi\right)^2 - \frac{V'_4}{a^4} \delta\chi \left(\partial^4\delta\chi\right) \nb\\
&& - \frac{V_6}{a^6} \left(\partial^2\delta\chi\right)\left(\partial^4\delta\chi\right) + \frac{\bar{A}c'_1}{a^2} \delta\chi\left(\partial^2\delta\chi\right) \nb\\
&& + \bar{A}c_1 \Big[{\cal{P}}_1|_{(2)}\Big] + \frac{\bar{A}c_2}{a^2}\left(\partial_k\delta\chi\right)^2 \nb\\
&& + \frac{c_1 \delta A}{a^2}\left(\partial^2\delta\chi\right),
\eqn
\bqn
\lb{C3.8}
\mathscr{L}_{\chi}|^{\text{HL}}_{(3)} &=& 2 a^{-2} \left(\frac{1}{2} + V_1\right)\zeta\left(\partial_k\delta\chi\right)^2  - \frac{V'_1}{a^2} \delta\chi\left(\partial_k\delta\chi\right)^2 \nb\\
&& -2 \frac{V_2}{a^2} \left(\partial^2\delta\chi\right)\Big[{\cal{P}}_1|_{(2)}\Big] - \frac{V'_2}{a^4} \left(\delta\chi\right) \left(\partial^2\delta\chi\right)^2 \nb\\
&& - \frac{V_3}{a^6} \left(\partial^2\delta\chi\right)^3 - V_4 \Big[{\cal{P}}_2|_{(3)}\Big] \nb\\
&& - V'_4 \left(\delta\chi\right) \Big[{\cal{P}}_2|_{(2)}\Big] - \frac{V''_4}{2a^4}\left(\delta\chi\right)^2\left(\partial^4\delta\chi\right) \nb\\
&& - \frac{V_5}{a^6} \left(\partial^4\delta\chi\right)\left(\partial_k\delta\chi\right)^2 \nb\\
&& - \frac{V'_6}{a^6}\left(\delta\chi\right)\left(\partial^2\delta\chi\right)\left(\partial^4\delta\chi\right) \nb\\
&& - \frac{V_6}{a^6} \Big[(\partial^2\delta\chi) \left(a^4{\cal{P}}_2|_{(2)}\right)+ (\partial^4\delta\chi) \left(a^2{\cal{P}}_1|_{(2)}\right) \Big] \nb\\
&& + \bar{A} c_1 \Big[{\cal{P}}_1|_{(3)}\Big] + \bar{A} c'_1 \left(\delta\chi\right)\Big[{\cal{P}}_1|_{(2)}\Big]\nb\\
&& + \frac{\bar{A}}{a^2}\frac{c''_1}{2} \left(\delta\chi\right)^2\left(\partial^2\chi\right)\nb\\
&& + \frac{\bar{A}}{a^2} \Big[-2c_2\zeta + c'_2\delta\chi \Big] \left(\partial_k\delta\chi\right)^2 \nb\\
&& + \frac{\delta A}{a^2} \Big[c'_1 \delta\chi \left(\partial^2\delta\chi\right) + c_1a^2  {\cal{P}}_1|_{(2)} \nb\\
&& ~~~~~~~~ + c_2 \left(\partial_k\delta\chi\right)^2 \Big],
\eqn
where
\bqn
\lb{C3.9}
a^2{\cal{P}}_1|_{(2)} &=&\left(\partial_k \zeta\right)\left(\partial^k \delta \chi\right) - 2 \zeta \left( \partial^2 \delta \chi\right),\nb\\
\lb{C3.10}
a^4 {\cal{P}}_2|_{(2)} &=&\Big[\left(\partial_k\delta\chi\right)\left(\partial^k\partial^2\zeta\right) + 2\left(\partial_k\zeta\right)\left(\partial^k\partial^2\delta\chi\right)\Big] \nb\\
&& + 2\left(\partial_{ij}\zeta\right)\left(\partial^{ij}\delta\chi\right) - 4\zeta \left(\partial^4\delta\chi\right),\nb\\
\lb{C3.12}
a^2{\cal{P}}_1|_{(3)}&=& 3\zeta^2 \partial^2\delta\chi,\nb\\
\lb{C3.13}
a^4 {\cal{P}}_2|_{(3)}&=& 8\zeta^2\partial^4\delta\chi - 8 \zeta \left(\partial_{ij}\zeta\right)\left(\partial^{ij}\delta\chi\right) \nb\\
&& -4\zeta \Big[\left(\partial_k\delta\chi\right)\left(\partial^k\partial^2\zeta\right) + 2\left(\partial_k\zeta\right)\left(\partial^k\partial^2\delta\chi\right)\Big] \nb\\ 
&& + \left(\partial^i\zeta\right)\Big[\left(\partial^j\zeta\right)\left(\partial_{ij}\delta\chi\right) + \left(\partial^j\delta\chi\right)\left(\partial_{ij}\zeta\right)\Big].\nb\\
\eqn

One can further substitute $\delta\chi,B$ and $\delta A$ in terms of $\zeta$ into  Eqs. (\ref{b20.a})-(\ref{b20.c}). 
See \cite{Collins} for a very good and detailed review of the subject.

\section*{Appendix C:  Determination of the integration constants}
 \renewcommand{\theequation}{C.\arabic{equation}} \setcounter{equation}{0}
We find that in the original work \cite{TransPl}, one of $B_1$ and $B_2$ must be zero in the case of the Corley-Jacobson dispersion with $b_m > 0$.
In fact, considering  their equations (134) and (135), we find that  the terms in the parentheses 
\bqn
\lb{D.1}
B_1 &\propto& \left(\text{cos}~y_1 - i~\text{sin}~y_1 - e^{\mp i \pi \nu} \text{cos}~x_1 + ie^{\mp i \pi \nu} \text{sin}~x_1\right) \nb\\
	&=& e^{-iy_1} - e^{\mp i \pi \nu}e^{-ix_1}= e^{-iy_1}\left(1-e^{\mp i \pi \nu}e^{- i \pi \nu}\right), \nb\\
B_2 &\propto& \left(\text{cos}~y_1 + i~\text{sin}~y_1 - e^{\mp i \pi \nu} \text{cos}~x_1 - ie^{\mp i \pi \nu} \text{sin}~x_1\right) \nb\\
	&=& e^{iy_1} - e^{\mp i \pi \nu}e^{ix_1}= e^{iy_1}\left(1-e^{\mp i \pi \nu}e^{+ i \pi \nu}\right),
\eqn
where we have used the definition of $x$ and $y$ in equation (128), $x=y+\pi\nu$. One immediately sees that either $B_1\neq 0, B_2 = 0$ (upper signs, no excitation and only ``positive frequencies" exist), or $B_1= 0, B_2 \neq 0$ (lower signs, complete excitation and only ``negative frequencies" exist).
Similar problems occur for their $A_1$ and $A_2$ too (take the large argument asymptotes of the Bessell functions). We believe this is an artifact of the approximation procedure, and that in general, both the ``positive-" and ``negative-frequency" modes should appear and the magnitude of the ``negative-frequency" modes should be limited by considerations from the back-reaction problem \cite{back, Holman}.

Below we re-derive the coefficients following the BM guidelines of matching, but within our concrete model.
From (\ref{b31}), the mode function $r_k(t)$ satisfies the  EoM
\bqn
\lb{D.05}
\ddot{r}_k(t) + \left(3H + m^2 + \frac{\omega^2_k}{a^2} \right) r_k(t) = 0,
\eqn
where $m^2 = \beta_4/\beta_0$, $\omega^2_k$ is the same as defined in Eq. (\ref{b35}) and a dot represents differential w.r.t time $t$. In Region I where the nonlinear effects in the dispersion dominates, the solution takes the asymptotic form (below we use $t$ and conformal time $\eta$ interchangeably, noting that $a d\eta = d t.$), 
\bqn
\lb{D.06a}
r^{\text{I}}_k(t) &\simeq& A_1 \exp\left[-i \frac{H^2 k^3}{3 M^2}\Big(\eta^3-\eta^3_i\Big)\right] \nb\\
&& + A_2  \exp\left[+i \frac{H^2 k^3}{3M^2}\Big(\eta^3-\eta^3_i\Big)\right]\nb\\
&=& \tilde{A}_1 \exp\left[-i \frac{H^2 k^3}{3 M^2}\eta^3\right] + \tilde{A}_2 \exp\left[+i \frac{H^2 k^3}{3M^2}\eta^3\right],\nb\\
\tilde{A}_1 &\equiv& A_1 \exp\left[+i \frac{H^2 k^3}{3M^2}\eta^3_i\right],\nb\\
\tilde{A}_2 &\equiv& A_2 \exp\left[-i \frac{H^2 k^3}{3M^2}\eta^3_i\right], \\
\lb{D.06b}
M^{-4} &\equiv& \frac{\beta_3a^4}{\beta_0}= \frac{2\left|V_6 + 2\pi G c^2_1 \alpha_2\right|}{f +4 \pi G c^2_1 / |c^2_{\psi}|}.
\eqn
It can be shown that $M$ here is related to $M_*$ defined in \cite{Inflation} through $M^4_* = b_3 M^4$. Since $b_3 \sim {\cal{O}}(1)$,
 the condition $H^2/M^2_* \ll 1$ implies  $H^2/M^2 \ll 1$.

In Region II, the mode function  is a superposition of plane waves, 
\bqn
\lb{D.07a}
r^{\text{II}}_k(t) &\simeq& B_1 \exp\left[-i c_s k\left(\eta-\eta_1\right)\right] + B_2 \exp\left[i c_s k\left(\eta-\eta_1\right)\right]\nb\\
&=& \tilde{B}_1 \exp\left[-i c_s k\eta\right] + \tilde{B}_2 \exp\left[+i c_s k\eta\right], \nb\\
\tilde{B}_1 &\equiv& B_1 \exp\left[+i c_s k\eta_1\right],\nb\\
\tilde{B}_2 &\equiv& B_2 \exp\left[-i c_s k\eta_1\right], \\
\lb{D.07b}
c^2_s &\equiv& \frac{\beta_1}{\beta_0} = \frac{1+ 2 V_1 + 2 \bar{A} (c_1' - c_2) - 4\pi G c^2_1 (1-\bar{A})}{f +4 \pi G c^2_1 / |c^2_{\psi}|}.\nb\\
\eqn
The initial conditions  are chosen  such that  
\bqn
\lb{D.10}
r_k(t_i) = \frac{v_k(\eta_i)}{z(\eta_i)}=\frac{1}{z(\eta_i)\sqrt{2\omega(\eta_i)}},
\eqn
whereas its initial time-derivative takes value such that the energy density is minimized at the initial time (not necessarily infinite past $\eta_i \to - \infty$)
\bqn
\lb{D.11}
r'_k(t_i) = \Big(\frac{v_k(\eta_i)}{z(\eta_i)}\Big)'\nb\\
	&=& \pm i \frac{1}{z(\eta_i)}\sqrt{\frac{\omega(\eta_i)}{2}}.
\eqn
At the time of matching between region I and II, $t_1$, 
\bqn
\lb{D.12}
r^{\text{I}}_k(t_1) = r^{\text{II}}_k(t_1), ~~~~~ r'^{\text{I}}_k(t_1) = r'^{\text{II}}_k(t_1).
\eqn
These four conditions (\ref{D.10})$\sim$(\ref{D.12}) fix the four undetermined constants
\bqn
\lb{D.20a}
\tilde{A}_{1+} &=& \frac{\exp\left[+i \frac{H^2 k^3}{3M_*^2}\eta^3_i\right]}{z\left(\eta_i\right)\sqrt{2\omega\left(\eta_i\right)}} \Big(1 + \frac{2i}{\eta_i \omega\left(\eta_i\right)}\Big), \nb\\
\tilde{A}_{2+} &=& \frac{\exp\left[-i \frac{H^2 k^3}{3M_*^2}\eta^3_i\right]}{z\left(\eta_i\right)\sqrt{2\omega\left(\eta_i\right)}} \Big(0 - \frac{2i}{\eta_i \omega\left(\eta_i\right)}\Big), \nb\\
\omega \left(\eta_i\right) &\simeq& \frac{k^3 H^2 \eta^2_i}{M_*^2},
\eqn
for $r'_k(t_i) = + i \frac{1}{z(\eta_i)}\sqrt{\frac{\omega(\eta_i)}{2}}$ (positive frequency choice), or 
\bqn
\lb{D.20b}
\tilde{A}_{1-} &=& \frac{\exp\left[+i \frac{H^2 k^3}{3M_*^2}\eta^3_i\right]}{z\left(\eta_i\right)\sqrt{2\omega\left(\eta_i\right)}} \Big(0 + \frac{2i}{\eta_i \omega\left(\eta_i\right)}\Big), \nb\\
\tilde{A}_{2-} &=& \frac{\exp\left[-i \frac{H^2 k^3}{3M_*^2}\eta^3_i\right]}{z\left(\eta_i\right)\sqrt{2\omega\left(\eta_i\right)}} \Big(-1 - \frac{2i}{\eta_i \omega\left(\eta_i\right)}\Big).
\eqn
for $r'_k(t_i) = - i \frac{1}{z(\eta_i)}\sqrt{\frac{\omega(\eta_i)}{2}}$ (negative frequency choice). And
\bqn
\lb{D.20c}
\tilde{B}_1 &=& \frac{e^{+ic_sk\eta_1}}{2} \Bigg[\left(1+\frac{k^2H^2\eta^2_1}{c_sM_*^2}\right)\tilde{A}_1 \exp\left(-i \frac{H^2 k^3}{3M_*^2}\eta^3_1\right)\nb\\
		&&~~~  + \left(1-\frac{k^2H^2\eta^2_1}{c_sM_*^2}\right)\tilde{A}_2 \exp\left(+i \frac{H^2 k^3}{3M_*^2}\eta^3_1\right)\Bigg], \nb\\
\tilde{B}_2 &=& \frac{e^{-ic_sk\eta_1}}{2} \Bigg[\left(1-\frac{k^2H^2\eta^2_1}{c_sM_*^2}\right)\tilde{A}_1 \exp\left(-i \frac{H^2 k^3}{3M_*^2}\eta^3_1\right)\nb\\
		&&~~~  + \left(1+\frac{k^2H^2\eta^2_1}{c_sM_*^2}\right)\tilde{A}_2 \exp\left(+i \frac{H^2 k^3}{3M_*^2}\eta^3_1\right)\Bigg]. \nb\\
\eqn
The time of matching $\eta_1$ (or equivalently, $t_1$) is a critical quantity, here we choose it to be \cite{Match}
\bqn
\lb{D.30a}
&&\omega^2_k (\eta_1) - c^2_s k^2 = c^2_s k^2,
\eqn
or from (\ref{b35})
\bqn
\lb{D.30b}
\beta_3\left(\eta_1\right) k^4 + \beta_2\left(\eta_1\right) k^2 = \beta_1.
\eqn
Now recall that by dividing the evolution history into three regions, we implicitly assumed  that the $\beta_2$ term was never dominant during the  history of inflation, as a result, the above condition can be approximated with
\bq
\lb{D.30c}
\beta_3\left(\eta_1\right) k^4 = \beta_1,
\eq
which leads to
\bq
\lb{D.31}
2 \left|V_6 + 2\pi G c^2_1 \alpha_2\right|k^4 H^4  \eta^4_1 = \beta_1.
\eq
Hence
\bqn
\lb{D.32}
\left(kH\eta_1\right)^2 &=& c_s M_*^2 \left(1+2 \delta_{\scriptscriptstyle{B}}\right),\nb\\
\eta_1 &\simeq& - \frac{\sqrt{c_s}}{k} \frac{M_*}{H} \left(1+ \delta_{\scriptscriptstyle{B}}\right),
\eqn
where we have used the definition of $c_s$ and $M_*$, and introduced a quantity $\delta_{\scriptscriptstyle{B}}$ to denote any deviations from (\ref{D.30c}), for example,  when we consider the $k^4$ term's minimal effect on $\eta_1$. (In \cite{Match}, it was found that a difference in this matching time could result in an unnecessary oscillatory component for the power spectrum, and we shall see below that this indeed happens through an extra oscillatory phase for the coefficients $B$.)

Looking at the equation (\ref{D.20c}), this choice of $\eta_1$ indicates that the influence of $\tilde{A}_2$ ($\tilde{A}_1$) on $\tilde{B}_1$ ($\tilde{B}_2$) is minimal. Also note that $\eta_i \omega\left(\eta_i\right) \sim ({H^2}/{M_*^2}) k^3\eta^3_i$ is in general a very large quantity, this allows us to make further approximations on the expressions of the coefficients $\tilde{A}$ and $\tilde{B}$ and arrive at
\bqn
\lb{D.35a}
\tilde{A}_{1+} &=& \frac{M_*}{h\sqrt{\beta_0}\sqrt{2k^3}} ~ e^{+i \phi_{\scriptscriptstyle{A}}}\left(1 + i \delta_{\scriptscriptstyle{A}}\right), \nb\\
\tilde{A}_{2+} &=& \frac{-M_*}{h\sqrt{\beta_0}\sqrt{2k^3}} ~ e^{-i \phi_{\scriptscriptstyle{A}}}\left(i \delta_{\scriptscriptstyle{A}}\right),\nb\\
\tilde{B}_{1+} &=& \frac{M_*}{h\sqrt{\beta_0}\sqrt{2k^3}} ~ e^{i \phi_{\scriptscriptstyle{B1}}}\left(1+ i \delta_{\scriptscriptstyle{A}} + \delta_{\scriptscriptstyle{B}} \right) , \nb\\
\tilde{B}_{2+} &=& \frac{-M_*}{h\sqrt{\beta_0}\sqrt{2k^3}}\left(i \delta{\scriptscriptstyle{A}}e^{-i \phi_{\scriptscriptstyle{B1}}} +\delta_{\scriptscriptstyle{B}}e^{i \phi_{\scriptscriptstyle{B2}}} \right), \nb\\
\eqn
or 
\bqn
\lb{D.35b}
\tilde{A}_{1-} &=& \frac{M_*}{h\sqrt{\beta_0}\sqrt{2k^3}} ~ e^{+i \phi_{\scriptscriptstyle{A}}}\left( i \delta_{\scriptscriptstyle{A}}\right), \nb\\
\tilde{A}_{2-} &=& \frac{-M_*}{h\sqrt{\beta_0}\sqrt{2k^3}} ~ e^{-i \phi_{\scriptscriptstyle{A}}}\left(1 + i \delta_{\scriptscriptstyle{A}}\right),\nb\\
\tilde{B}_{1-} &=& \frac{M_*}{h\sqrt{\beta_0}\sqrt{2k^3}} \left(i \delta{\scriptscriptstyle{A}}e^{i \phi_{\scriptscriptstyle{B1}}} +\delta_{\scriptscriptstyle{B}}e^{-i \phi_{\scriptscriptstyle{B2}}} \right),\nb\\
\tilde{B}_{2-} &=& \frac{-M_*}{h\sqrt{\beta_0}\sqrt{2k^3}} ~ e^{-i \phi_{\scriptscriptstyle{B1}}}\left(1+ i \delta_{\scriptscriptstyle{A}} + \delta_{\scriptscriptstyle{B}} \right),
\eqn
where 
\bqn
\lb{D.35c}
\delta_{\scriptscriptstyle{A}} &\equiv& \frac{2}{\eta_i \omega\left(\eta_i\right)} = \frac{2M_*^2}{k^3\eta^3_iH^2}, \\ 
\phi_{\scriptscriptstyle{A}} &\equiv& \frac{H^2 k^3}{3M_*^2}\eta^3_i, \nb\\
\lb{D.35d}
\phi_{\scriptscriptstyle{B1}} &\equiv& \frac{c_s k \eta_1}{3} \left[\left(1+2\delta_{\scriptscriptstyle{B}}\right) \left(\frac{\eta_i}{\eta_1}\right)^3 + 2 \left(1-\delta_{\scriptscriptstyle{B}}\right)\right], \nb\\
\phi_{\scriptscriptstyle{B2}} &\equiv& \frac{c_s k \eta_1}{3} \left[\left(1+2\delta_{\scriptscriptstyle{B}}\right) \left(\frac{\eta_i}{\eta_1}\right)^3 - 2 \left(2+\delta_{\scriptscriptstyle{B}}\right)\right].\nb\\
\eqn


\end{document}